\documentclass{llncs}
\usepackage[T1]{fontenc}
\usepackage{graphicx}
\usepackage{subfig}
\usepackage{subcaption}
\usepackage{adjustbox}
\usepackage{hyperref}
\usepackage{amsmath}
\usepackage{comment}
\usepackage{tikz-network}
\usepackage{booktabs}
\usepackage{rotating}
\usepackage{longtable} 
\usepackage{pdflscape}   
\usepackage{xurl}
\usepackage{tabularx}

\begin{document}

\title{Multilayer Network Analysis of European
Regional Flows}
\titlerunning{Multilayer Network Analysis of European Regional Flows}

\author{Emanuele Calò
\orcidID{0009-0007-5545-907X} \and
Angelo Facchini
\orcidID{0000-0003-2652-5238}}

\authorrunning{E. Calò and A. Facchini}

\institute{IMT School for Advanced Studies Lucca, Piazza San Francesco 19, 55100 Lucca, Italy
}

\maketitle           
\begin{abstract}
In Regional Economics, the attractiveness of regions for capital, migrants, tourists, and other kinds of flows is a relevant topic. Usually, studies in this field explore single flows, characterizing the dimensions of territorial attractiveness separately, rarely considering the interwoven effect of flows. Here, we investigate attractiveness from a multi-dimensional perspective (i.e., dealing with different flows), asking how various types of regional flows collectively shape the attractiveness dynamics of European regions. We analyze eight distinct flow types across NUTS2 regions from 2010 to 2018, employing a multilayer network approach. Notably, the multilayer approach unveils insights that would be missed in single-layer analyses. Community detection reveals complex structures that demonstrate the cohesive power of national borders and the existence of strong cross-border ties in specific regions. Our study contributes to a more nuanced understanding of regional attractiveness, with implications for targeted policy interventions in regional development and European cohesion.

\end{abstract}

\section{Introduction}

The economic vitality and developmental trajectory of European regions are increasingly defined not by their intrinsic attributes alone, but by their position within a complex web of inter-regional connections. This perspective aligns with the foundational concept of a \emph{space of flows}~\cite{castells1996information}, where the movement of capital, people, goods, and knowledge constitutes the fundamental architecture of the contemporary economy. In this relational view~\cite{amin2004regions,bathelt2011relational}, the prosperity and functional role of a region are contingent on its connectivity, i.e., the strength and diversity of its ties to other regions. While this phenomenon is often studied in the context of territorial attractiveness~\cite{servillo2012territorial,russo2013attreg}---a crucial concept for integrating regional development strategies with the overarching goal of territorial cohesion~\cite{faludi2006european}---we argue that considering the underlying network dynamics is essential for a complete and nuanced understanding of regional development and prosperity in today's integrated~economy.

Traditionally, research has attempted to map these interdependencies through two main lenses. The first analyzes individual flows in isolation, such as tourism~\cite{cracolici2009attractiveness}, migration~\cite{waltert2010landscape}, or investment~\cite{jackson1995attractiveness}. These analyses, while valuable, may provide an incomplete view that is unable to capture how different types of flows interact with one another. The second approach seeks a multidimensional view by creating composite indicators of territorial attractiveness~\cite{musolino2024new,OECD}. However, these methods primarily treat regions as independent entities, aggregating their internal characteristics while largely overlooking the network structure of the flows that connect them and define their functional roles within the wider European system. We argue that a fundamental gap exists in understanding how different regional flows come together to form a cohesive system of multidimensional connectivity.

This paper addresses this gap by proposing a paradigm shift from evaluating regional attractiveness as an inherent quality to analyzing multidimensional connectivity as a continuously interacting, relational process. We move beyond asking ``How attractive is a region?'' to asking ``What is the structural role of a region within the interconnected European system?''. To do this, we operationalize our framework using multilayer network science~\cite{kivela2014multilayer}, modeling the European space as a multilayer network where each layer represents a distinct type of flow. This allows us to re-interpret what has been termed \emph{revealed attractiveness}
~\cite{musolino2016attrattivita}—not as a simple sum of inflows, but as a region's emergent structural position resulting from the complex interplay across multiple, co-existing networks. Different spatial scales can be used for this analysis, ranging from the macro-scale (e.g., countries~\cite{reiner2017urban}) to the micro-geographical scale (e.g., neighborhoods~\cite{oner2017retail} or cities~\cite{lee2016conceptualization}). Regions at the NUTS2 scale provide an ideal compromise, being neither too large to hide local dynamics nor too small to miss broader patterns of inter-regional connection.

Our analysis of eight distinct flow types across European NUTS2 regions from 2010 to 2018 is guided by the following research questions:
\begin{itemize}
\item[R1] How does a region's importance transform when moving from an analysis of isolated flow types to a comprehensive multidimensional connectivity framework?
\item[R2] What emergent functional relationships and regional clusters are revealed by the interplay of different flow types in a multilayer framework?
\end{itemize}

Our study first addresses the conceptual need to better capture the complex, interconnected nature of European regions by developing a multidimensional connectivity framework. This framework advances knowledge by moving beyond traditional single-dimension models, allowing for a richer understanding of how regions interact across flows of capital, knowledge, and people. Operationally, we show that applying advanced multilayer network techniques (i.e., multiplex PageRank centrality~\cite{page1999pagerank} and Infomap community detection~\cite{rosvall2009map}) offers clear advantages over previous methods, as these tools can reveal intricate regional structures and patterns that traditional economic or econometric methods may overlook~\cite{FinalReport,Komornicki2023}. Empirically, we demonstrate the value of this approach by providing a structurally aware understanding of European regional dynamics that identifies key hubs, tracks changes in regional roles over time, and uncovers functional ties that transcend national borders. Together, this conceptual, methodological, and empirical integration provides new insights into the evolving fabric of regional connectivity in Europe.

Our analysis reveals that European regional networks are characterized by a core--periphery structure, where a few regions dominate connectivity across multiple flow types. The multilayer perspective provides a more nuanced view of regional importance than single-layer analyses, highlighting the significant enhancement in the importance of regions like Bratislava and Leipzig. Furthermore, our community detection algorithm uncovers robust regional clusters, confirming the cohesive power of national borders but also revealing strong, unexpected cross-border functional regions. These findings underscore the necessity of a relational, multidimensional perspective for policymakers aiming to foster balanced regional development. The remainder of this paper details the data and methodology used, presents the full results of our network analysis, and discusses the implications of our findings for regional science and European policy.

\section{Materials and Methods}\label{data}
\subsection{Data}
ESPON is an EU-funded program providing territorial analyses, data, and maps. 
The dataset utilized in this study is derived from the IRiE ESPON project~\cite{IRiE} and includes region-to-region (NUTS 2 level, 2016 version) origin--destination (OD) matrices covering various domains such as People Tourism, People Migration, Freight of Goods by transport mode, Capital Foreign Direct Investment (FDI), Knowledge (Erasmus students), People Passengers by transport mode, Capital Remittances, and Knowledge (Horizon 2020).
Combining these datasets enables us to create a comprehensive, multidimensional view of regional connectivity, capturing how distinct flows of people, capital, goods, and knowledge collectively define the structure of the European space.
This choice of layers is not arbitrary but is instead grounded in a substantial body of literature that identifies these specific flows as complementary channels of economic and social integration. Diaspora networks are known to reduce information costs and boost trade~\cite{rauch2002ethnic,peri2010trade,docquier2010skilled}; tourism has pro-trade effects~\cite{santana2011tourism}; passenger and air connectivity enable knowledge transfer and are associated with FDI~\cite{ishutkina2008analysis,fageda2017international}; and remittances are a primary measure of diasporic and financial links~\cite{WorldBankRemittances2023}. Therefore, combining these distinct but interrelated flows in a multilayer network is an appropriate and well-established method for capturing a holistic view of connectivity~\cite{kivela2014multilayer,boccaletti2014structure,bonaccorsi2019country}.

The data encompass the flows between 297 European regions recorded annually. Different periods are covered for each flow type, i.e., 2010--2014 for Erasmus, 2015--2020 for Horizon 2020, and 2010--2018 for all other categories.
Distinct methodologies were employed by the researchers who collected and processed each type of OD dataset. They gathered and harmonized various data sources at both European and national levels, initially focusing on country-to-country flows. While most of the data were raw, some flows were estimated using specific techniques (for further technical details, refer to the online documentation). These country-level flows were then decomposed to the regional level for more detailed analysis. In Table \ref{tab:flow_overview}, we present an overview of the data used in this study. It is important to note that the column \``Methodology''
 describes the procedures employed by the original data collectors, while the column ``Our Analysis'' outlines the additional steps we performed for our specific analysis.

\subsection{A Network Science Framework}

To address our research questions, we employ network science~\cite{newman2018networks}, which offers effective tools for analyzing complex systems of interconnected entities, making it particularly well-suited for studying the multifaceted nature of regional connectivity in Europe. 
This field has demonstrated that complex real-world systems---from biological to economic networks, as well as spatial networks~\cite{barthelemy2011spatial}---exhibit significant and interpretable non-random structures~\cite{albert2002statistical,cimini2019statistical}.
Specifically, multilayer networks provide a powerful framework for modeling complex systems where entities are connected through multiple types of relationships simultaneously ~\cite{bianconi2018multilayer}. These networks, consisting of multiple interrelated layers interacting with each other, can encompass various domains such as social networks (with layers for friendship, family, and professional ties), financial markets (with layers for different asset classes), and multimodal transportation systems. The multilayer structure significantly influences the dynamics within these systems, often leading to unexpected behaviors. For example, diffusion on multilayer transportation networks can significantly speed up with respect to diffusion on single layers~\cite{gomez2013diffusion}. This is because the ability to switch between layers—like a passenger switching from a metro to a bus—creates new pathways that would be invisible in separate analyses of each transportation mode. More broadly, this perspective reveals hidden structural correlations, identifies system-wide vulnerabilities, and allows for a more realistic modeling of contagion or information spreading processes that simultaneously leverage different types of connections~\cite{boccaletti2014structure}.

In the context of this study, the value of the multiplex perspective is that it allows us to ask questions that are inaccessible from a single-layer viewpoint. 
While previous studies have collected and examined territorial flow data~\cite{kang2020multiscale} and, in some cases, employed network science~\cite{provenzano2018mobility}, with our approach, we move beyond single-flow analyses by integrating multiple flow types into a comprehensive multilayer network, providing a more holistic and structurally aware view of regional interactions.
Our approach, therefore, is not merely exploratory; it is a structured investigation designed to test for patterns within the European regional system.
Multilayer centrality measures, for instance, identify regions that may not be dominant in any single flow type but are crucial connectors when all flows are considered together, acting as versatile hubs in the overall European network. Furthermore, the multilayer approach fundamentally enhances our ability to detect meaningful functional communities. Instead of finding clusters based solely on tourism or investment, our analysis identifies groups of regions that are strongly interconnected across a combination of people, capital, and knowledge flows. This reveals cohesive economic and social blocks that are defined by their multifaceted relationships, offering a more robust picture of regional integration than any single layer could provide. This methodical progression from single-layer to multilayer analysis, and from basic structural properties to complex community structures, allows us not only to comprehensively map the multifaceted nature of regional interconnectedness in Europe but also contributes to the broader field of network science by demonstrating its applicability and value in regional studies and policy analysis.\\

\subsection{Single-Layer Network Construction}

As a first step, we represent each of the eight flow types as a separate weighted directed network (a ``layer''). In each layer, nodes correspond to regions, links represent flows between regions within a given year, and link weights denote the magnitude of these flows. Detailed information on the number of nodes, links, and density for each layer and year can be found in the Supplementary Information (Section \ref{s3.1}).

\subsection{Multilayer Network Construction}

\begin{figure}[h] 
    \centering
    \includegraphics[width=0.345\textwidth]{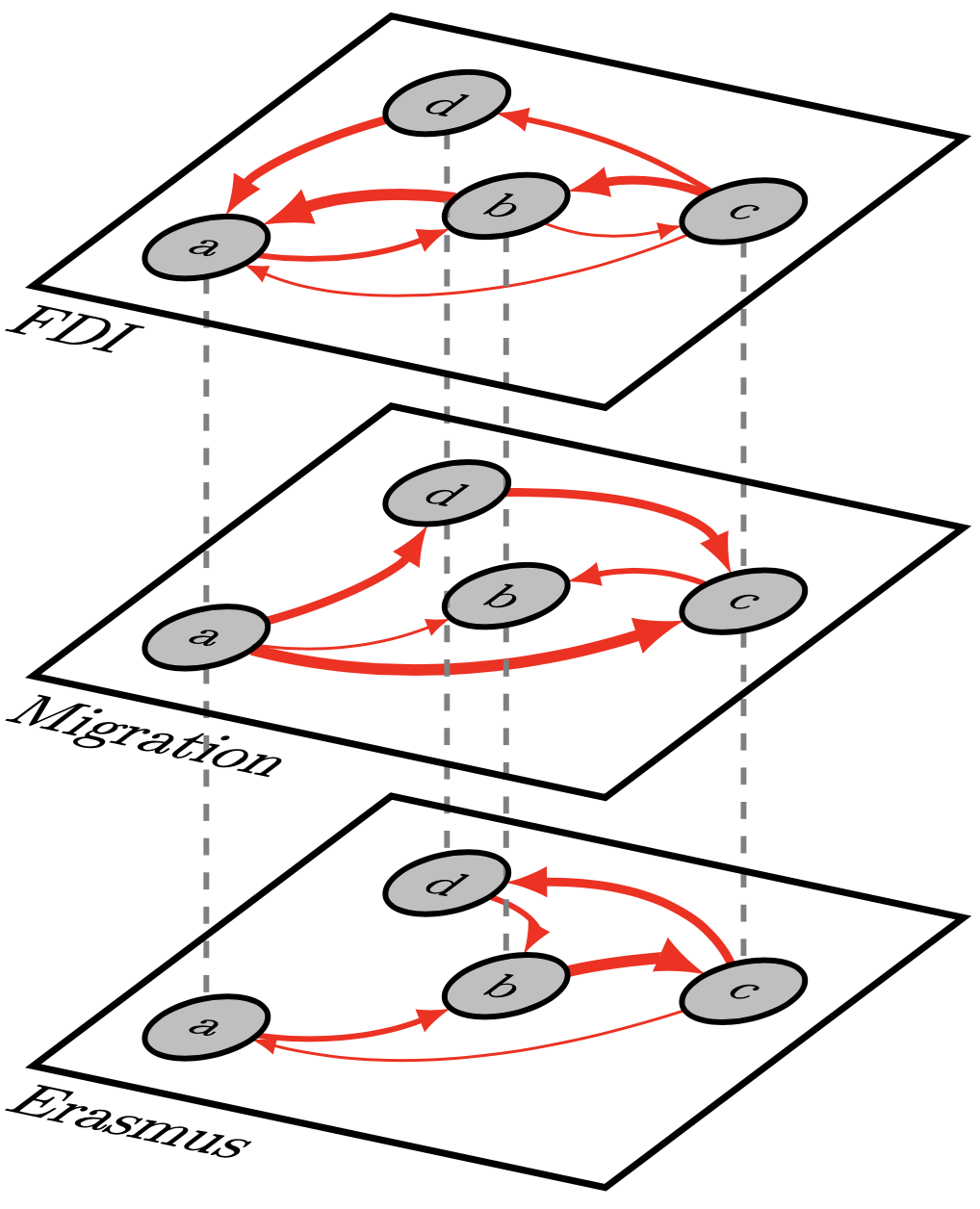}
    \caption{Illustrative schematic of a multiplex network. This conceptual diagram shows a network with three layers to represent different flow types. The nodes represent four hypothetical regions (a, b, c, and d), which are the same across all layers. The links are also hypothetical and are arranged differently on each layer for illustrative purposes, demonstrating how a region's connectivity can vary across dimensions.}
    \label{fig:multilayer_network}
\end{figure}

To analyze the complex interactions between different types of flows, we construct a multilayer network for each year by integrating all single-flow layers. Since the same set of regions is present across all layers, with connections existing only within each layer (i.e., no direct links between a tourism node and an FDI node), the resulting structure is a multiplex network~\cite{bianconi2018multilayer}. Figure \ref{fig:multilayer_network} provides a simplified, conceptual illustration of this multiplex structure. The diagram uses three layers for visual clarity and employs generic link patterns to demonstrate the concept; it is not a direct representation of our empirical data, which encompass all available flow types.
\vspace{-8pt}

\subsection{Analytical Techniques}
\subsubsection{Network Properties}

We focus initially on single layers by analyzing first-order properties, such as the strength distribution, which helps us understand the basic structure of the networks by examining how strongly regions are connected. Studying first-order properties is important because it provides fundamental insights into the connectivity and flow patterns within the network.
The out-strength ($s_i^{out}$) and in-strength ($s_i^{in}$) of node $i$ are defined as follows:
\begin{equation}
    s_i^{out} = \sum_jw_{ij}\quad \textup{and} \quad s_i^{in} = \sum_jw_{ji}
\end{equation}
where $w_{ij}$ represents the weight of the directed edge from node $i$ to node $j$. We examine the complementary cumulative distribution function (CCDF) for in-strength, out-strength, and total strength.

Next, we move on to second-order properties, such as assortativity, which measures the tendency of nodes to connect to others that are similar or different in some way. Specifically, we focus on the Weighted Average Nearest Neighbors Strength (WANNS), i.e., $\textup{WANNS}^{in,out}$, which is defined as follows:
\begin{equation}
    \textup{WANNS}_i^{in,out} =\frac{\sum_jw_{ij}s_j^{out}}{s_i^{in}}, \quad \forall i.
\end{equation}

This calculates the weighted mean of the strengths of a node's neighbors.

\subsubsection{Null Model: CReMA}

We make use of null models for validating the WANNS outcomes, adding a robust statistical foundation to our findings and enhancing the reliability and interpretability of the observed network structures. By comparing assortativity to a null model, we can discern whether the observed connections are due to underlying structural patterns or are random. This step is essential as it reveals deeper relational dynamics within the networks, better highlighting those topological aspects that are not immediately captured from first-order analysis. To achieve this, we compare our results with the CReMA null model~\cite{parisi2020faster}. This model reconstructs the network topology and assigns weights to established links by maximizing entropy under given constraints. As long as these constraints are met (on average), all possible configurations are equally likely. A specific instance of this model is the Directed Enhanced Configuration Model~\cite{squartini2015unbiased}, which constrains the sequences of in-degrees, out-degrees, in-strengths, and out-strengths. We use the NEMtropy package~\cite{vallarano2021fast} to solve the model, employing the Newton method for both binary and weighted reconstructions, as well as the dcm-exp model for binary reconstruction, to generate an ensemble.

\subsubsection{Centrality Measure: PageRank}

Next, we examine centrality measures, as these are vital for identifying the network's most important or influential nodes. They help us understand the roles different regions play in the network, whether key hubs or peripheral nodes. To quantify the importance of regions within single-layer networks, we compute the PageRank centrality measure~\cite{page1999pagerank}.

Once we have understood the basic one-dimensional features of the flow networks, we study the flows from a multi-dimensional perspective. To this aim, we study the multilayer PageRank centrality via the muxViz package (version 3.1) in R (version 4.4.0), ~\cite{de2015muxviz}. This comprehensive approach allows us to capture the complex interactions between different types of flows, providing a wide perspective of regional dynamics and their broader implications. Indeed, unlike its single-layer counterpart, multilayer PageRank accounts for a random walker that can move both within layers and jump between layers, thus capturing a region's influence not only within a specific flow type but across the entire integrated system.

\subsubsection{Community Detection: Infomap}

Finally, we apply community detection, using the Infomap algorithm, to the multilayer network, revealing clusters of regions across multiple types of flows. Specifically, we employ the Infomap algorithm~\cite{rosvall2009map}, which can detect hierarchical community structures within and across layers. Infomap optimizes a quality function related to the random walker's trajectory, revealing both broad and granular communities. Key parameters include the \emph{two-level}
setting for nested module detection and the multilayer relaxation rate for inter-layer movements. We conducted a sensitivity analysis on the relaxation rate to understand its impact on community detection. Detailed information on parameter selection and sensitivity analysis results are provided in Sections S2 and S3.3 in the Supplementary~Information.

\section{Results}\label{results}
\subsection{Single-Layer Networks}\label{s3.1}
\subsubsection{First-Order Properties}

\begin{figure}[t]
    \centering
    \includegraphics[width=\textwidth]{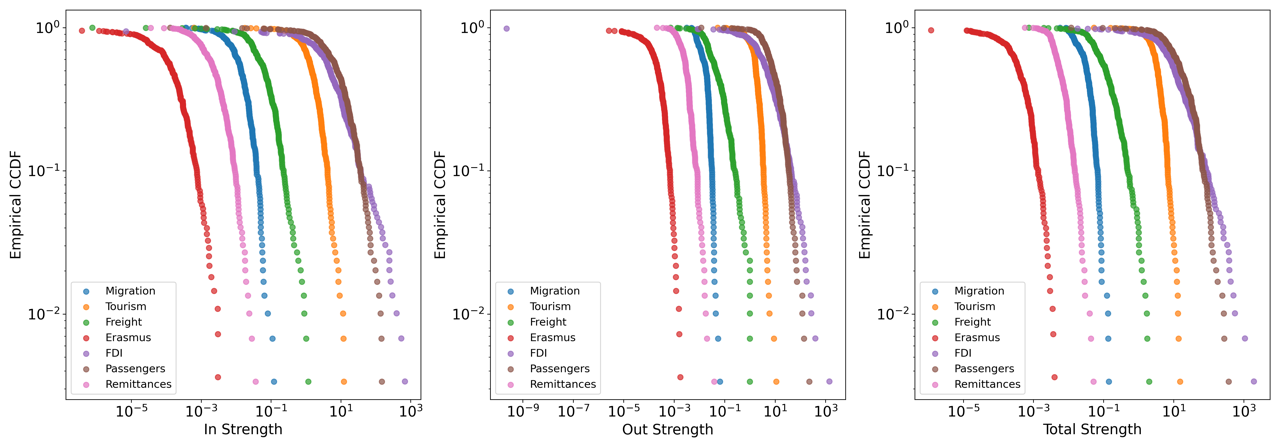}
    \caption{Complementary cumulative distribution function for the year 2010.}
        \label{fig:ccdf_2010}
\end{figure}

A key feature of many real-world networks is a heavy-tailed degree (or strength, in the weighted case) distribution. This general class of distributions signifies a consistent structural feature---the coexistence of a few nodes with exceptionally high connection strength and a large number of nodes with weak connectivity. To assess this pattern, we analyze the complementary cumulative distribution function (CCDF) of node strengths for the year 2010, as shown in Figure \ref{fig:ccdf_2010}. 
Indeed, all layers reveal a small number of regions acting as prominent hubs. Despite the somewhat limited range of these high-strength nodes, our results indicate that the networks exhibit this common characteristic of complex systems.
This pattern persists in the 2018 data, as demonstrated in the Supplementary Information, which includes further analyses depicting the relationship between in-strength and out-strength for all flow types in 2010 and 2018.

\subsubsection{Second-Order Properties}

A fundamental question in regional science is whether dominant regions preferentially interact with each other, reinforcing existing inequalities, or if they act as an integrative force, fostering connections with less-developed, peripheral areas. While the existence of hubs is a known phenomenon, we use assortativity analysis to gain further insights into the nature of their connectivity. This method allows us to distinguish between a network that is hub-dominated simply because hubs have many links and one that exhibits a specific, non-trivial connection pattern.
To isolate this effect, we compare our empirical findings against a robust null model. This model creates a precise counter-factual scenario---a network where each region retains its exact observed connectivity (both the number and strength of its links), but its connections are randomly rewired. This baseline shows the level of assortativity that would be present purely by chance, given the number and strength of each region's connections.

\begin{figure}[t]
    \centering
    \includegraphics[width=0.85\textwidth]{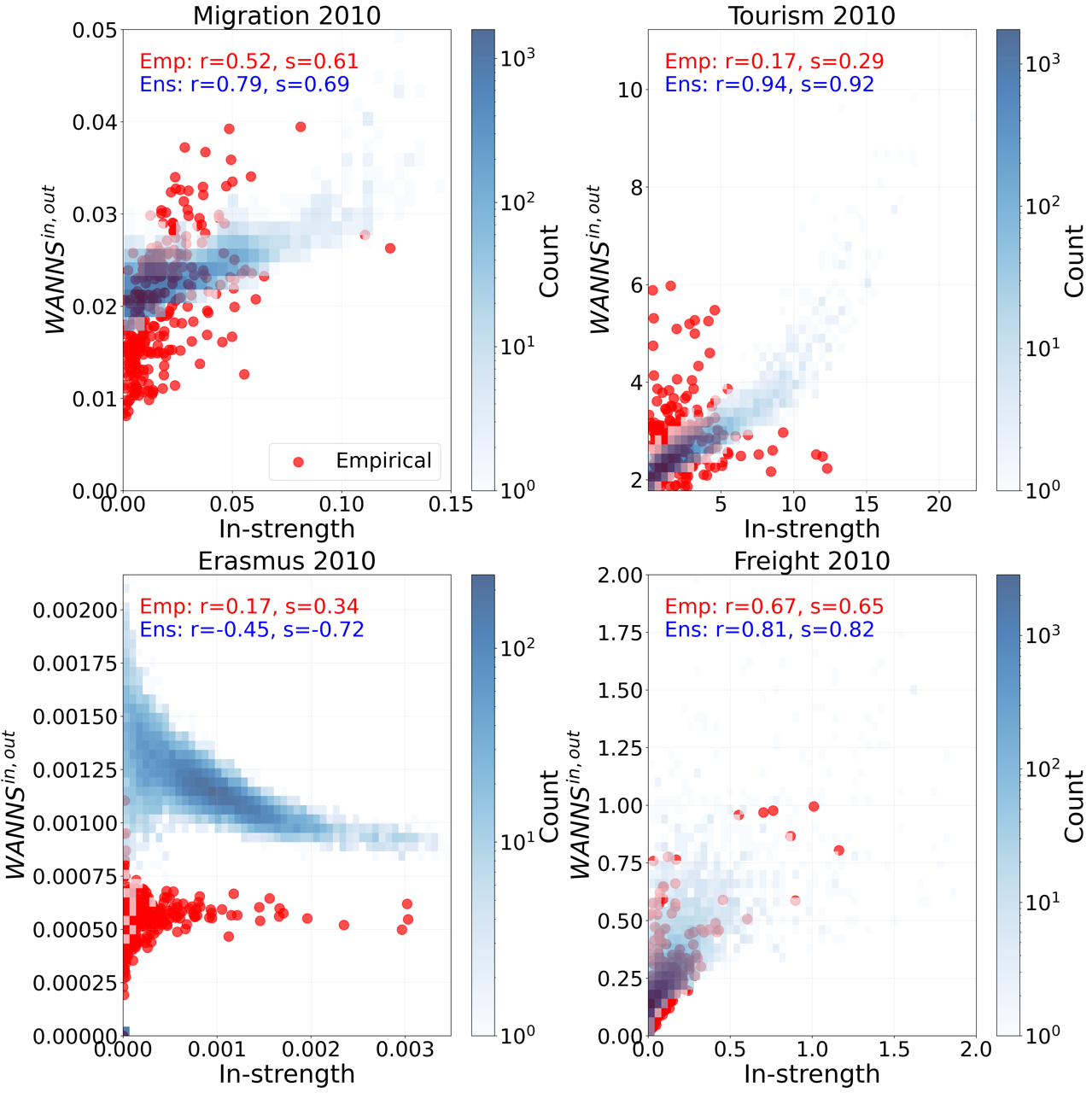}
    \caption{Empirical vs. ensemble WANNS for Migration, Tourism, Erasmus, and Freight (2010). Subplots show in-strength vs. $\textup{WANNS}^{in,out}$. Red: empirical data; blue histogram: ensemble distribution. Pearson's r and Spearman's s correlations provided.}
    \label{fig:wanns}
\end{figure}

Figure \ref{fig:wanns} presents the Weighted Average Nearest Neighbor Strength (WANNS) for the empirical networks, alongside 50 realizations drawn from the null model ensembles for Migration, Tourism, Erasmus, and Freight. In the Supplementary Information, we present the remaining three layers. The analysis reveals a predominantly assortative trend across all flow types. This assortative behavior indicates a positive correlation between node strengths and their neighbors' strengths, suggesting that regions with strong connectivity tend to connect with other strongly connected regions.
This analysis reveals the presence of a core--periphery structure, whereby the core is composed of regions that demonstrate a high level of interaction and exchange, while the periphery is constituted by regions with limited engagement in these flows.
Comparing the empirical results with the null model predictions, we observe that the null model consistently anticipates a stronger correlation between $\textup{WANNS}^{in,out}$ and in-strength. This pattern holds true for all flow types except for Erasmus, where the null model predicts negative correlation coefficients, and for passenger transport, where both Pearson's and Spearman's correlations are lower in the null model, deviating from the trend observed in other flow types. Similarly, for FDI, we note that Spearman's correlation coefficient of the null model is smaller than the empirical one.

The finding that our empirical networks are less assortative than expected by random chance is a significant and non-trivial result. It indicates that the European network is not as clustered as the distribution of its hubs would suggest. This finding challenges a simplistic view of a rigid core--periphery structure. It shows that hubs are also connecting to weaker, peripheral regions more often than predicted by the null model, suggesting a pattern of broader spatial integration and spillovers, despite the overall presence of a core--periphery structure.

Our approach provides a complementary perspective to methods, leveraging on exogenous inter-regional relationships that commonly use a pre-defined weighting matrix based on geographic distance or other contiguity measures~\cite{anselin1988spatial}. While in such methods, spatial spillovers are generally assumed on a geographical basis, here, we define relationships endogenously, highlighting the functional geography of the system as dictated by the flows themselves. The assortativity that we observe is therefore not only a function of geographic proximity but it also suggests that functionally central regions connect to other functionally central regions, regardless of how far apart they are. This allows us to uncover the true interaction topology of the European system, identifying non-local, long-distance corridors of interaction that a standard spatial econometric model would miss.

\subsection{Centrality Measure: PageRank}

\begin{figure}[t]
    \centering
    \includegraphics[width=0.85\textwidth]{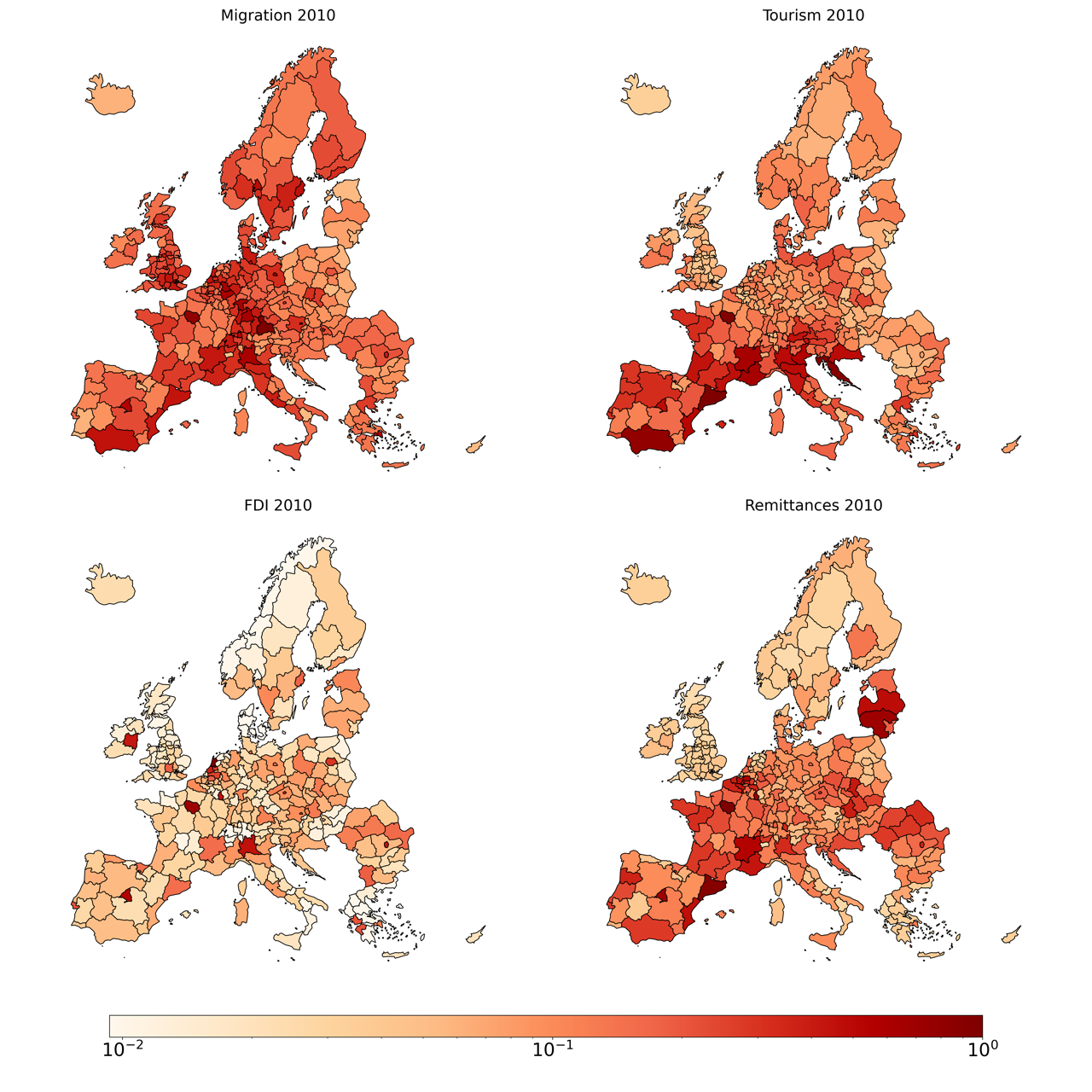}
    \caption{PageRank for Migration, Tourism, FDI, and Remittances in 2010. Colors are displayed on a logarithmic scale, with values normalized such that the region with the highest centrality is set to 1.}
    \label{fig:page_single_map_2010}
\end{figure}
Figure \ref{fig:page_single_map_2010} illustrates the spatial distribution of PageRank centrality values across European regions for Migration, Tourism, FDI, and Remittances in 2010. This visualization provides insights into the relative importance of regions within various flow networks, emphasizing the heterogeneity of regional centrality across different types of flows.
In the Supplementary Information, we present the remaining three layers and detailed tables showcasing the top 10 regions ranked by PageRank for all flow types in 2010. Moreover, we include further analyses on the relationship between PageRank and in-strength.

\begin{figure}[t]
    \centering
    \includegraphics[width=0.8\textwidth]{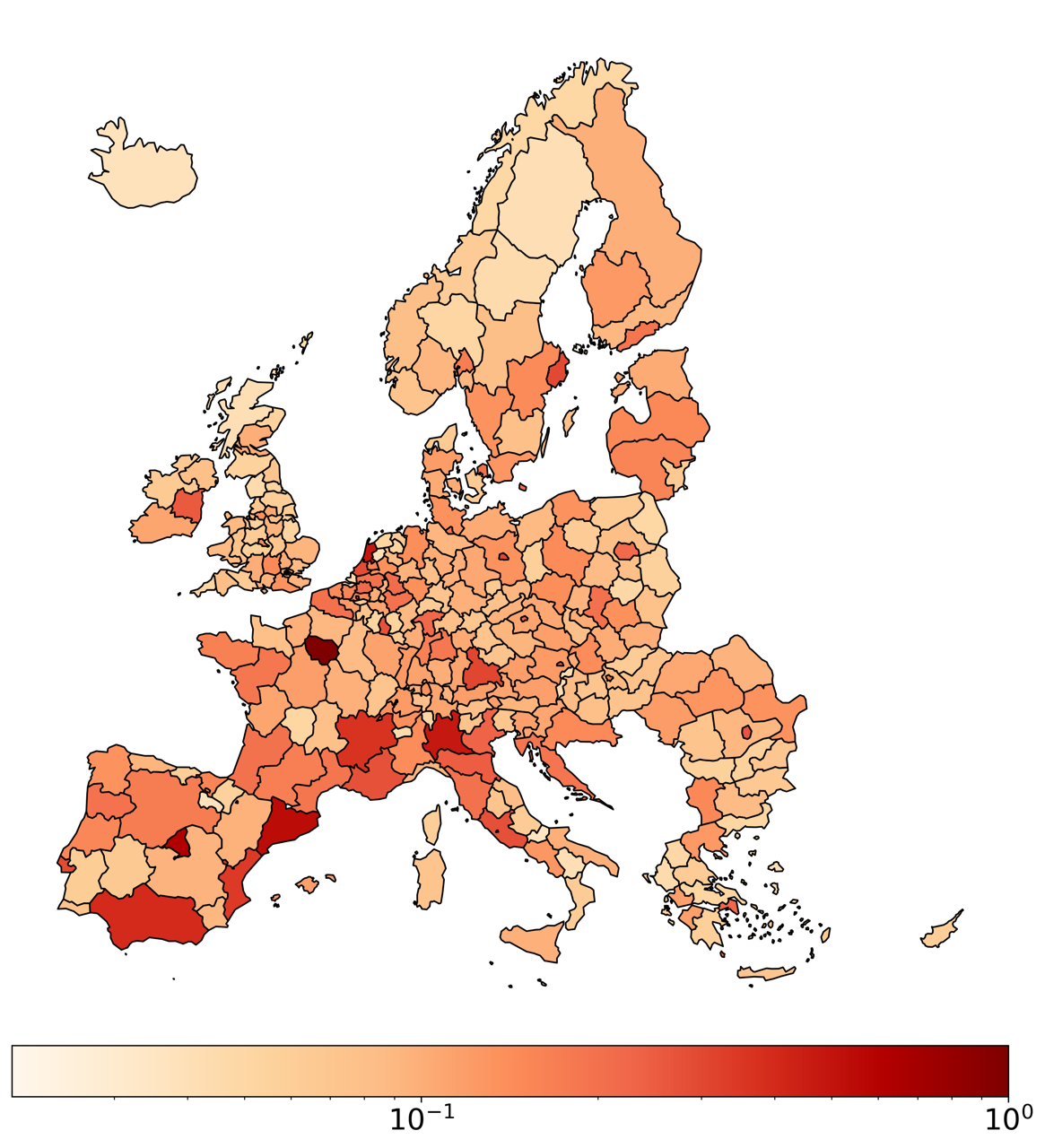}
    \caption{PageRank values for the 2010 multiplex. Colors are displayed on a logarithmic scale, with values normalized such that the region with the highest centrality is set to 1.}
    \label{fig:page_multi_map_2010}
\end{figure}

In contrast, Figure \ref{fig:page_multi_map_2010} depicts the spatial distribution of multiplex PageRank centrality values for the same year, offering a comprehensive view of regional significance within the interconnected multilayer network structure by integrating information from all flow~types.

\begin{figure}[t]
    \centering
    \includegraphics[width=\textwidth,height=\textheight,keepaspectratio]{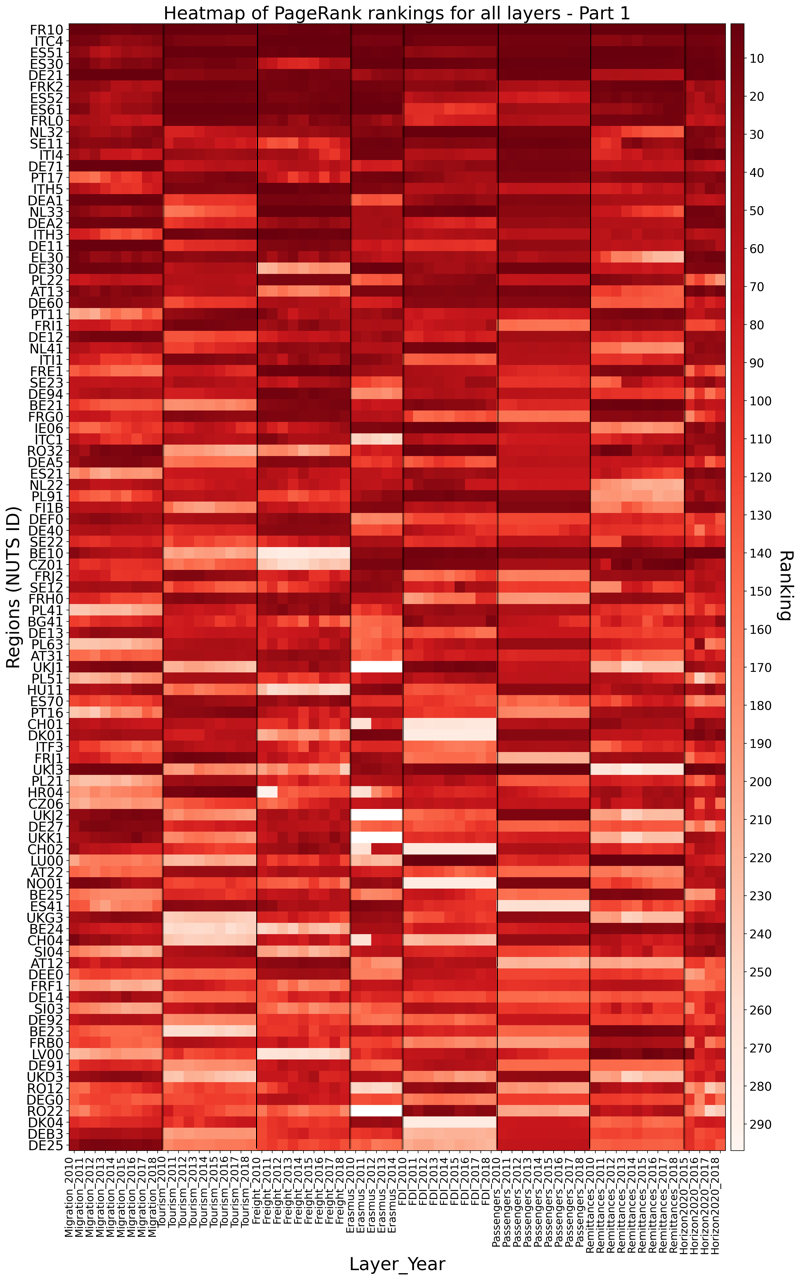}
    \caption{PageRank Heatmap - Top-Ranking regions across layers and years.}
        \label{fig:page_heat_1}
\end{figure}

\begin{figure}[t]
    \centering
    \includegraphics[width=\textwidth,height=\textheight,keepaspectratio]{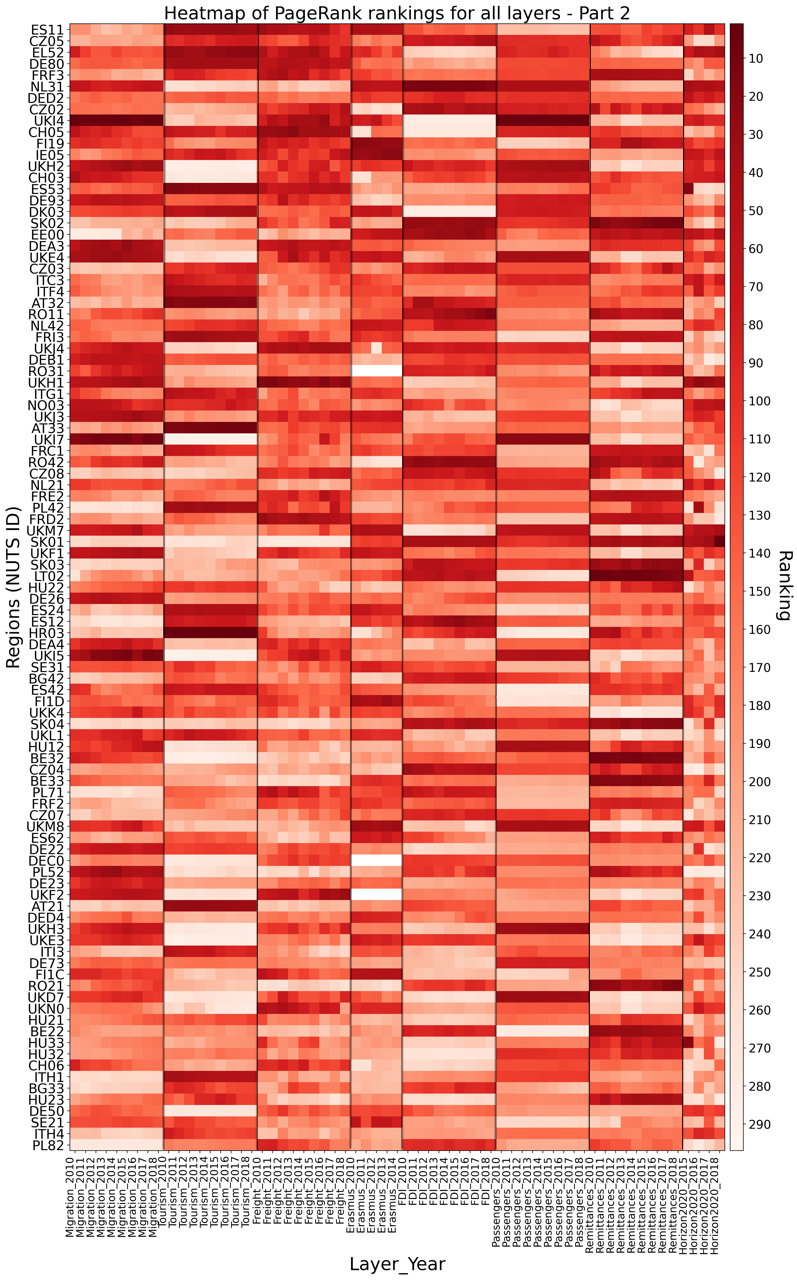}
    \caption{PageRank Heatmap - Mid-Ranking regions across layers and years.}
        \label{fig:page_heat_2}
\end{figure}

\begin{figure}[t]
    \centering
    \includegraphics[width=\textwidth,height=\textheight,keepaspectratio]{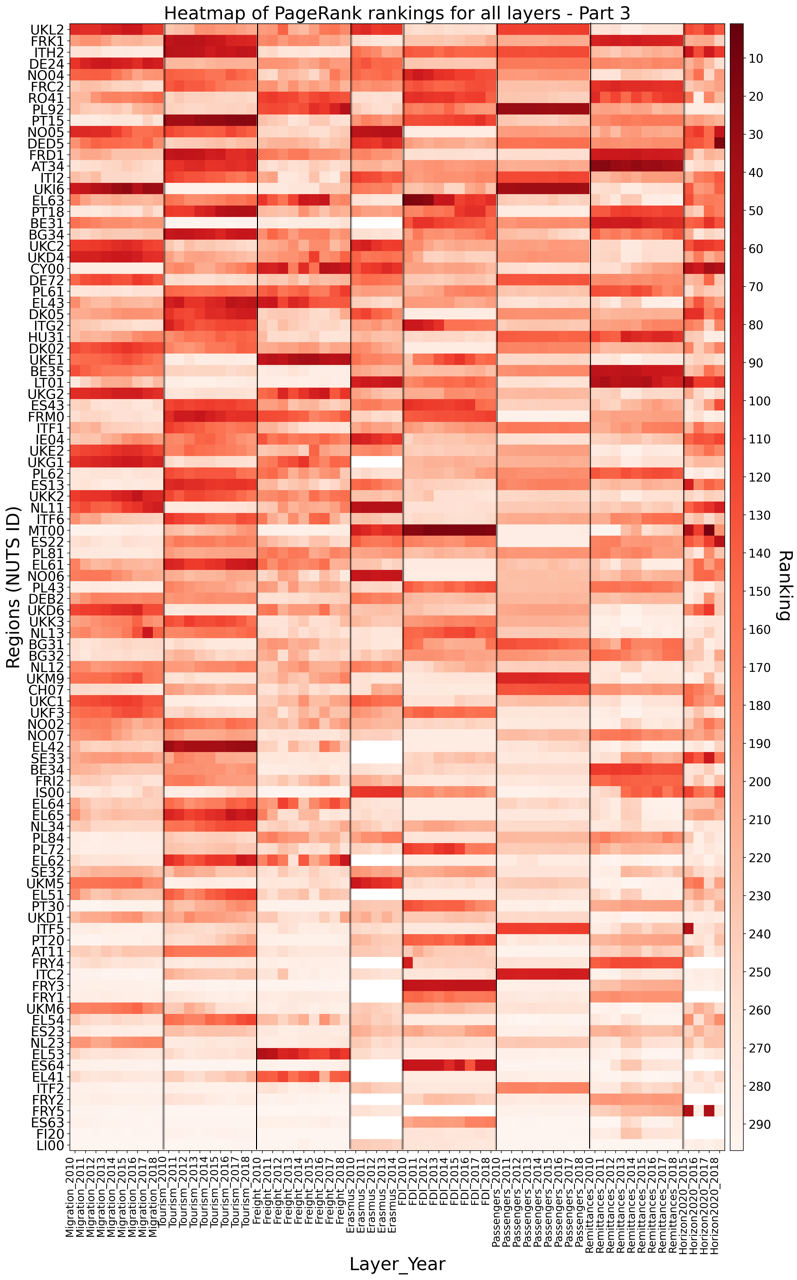}
    \caption{PageRank Heatmap - Bottom-Ranking regions across layers and years.}
        \label{fig:page_heat_3}
\end{figure}

To capture the temporal evolution of PageRank rankings, Figures \ref{fig:page_heat_1}--\ref{fig:page_heat_3} display a heatmap of these rankings for each layer, with regions ordered according to their average position across all layers and years. 
The figures present a series of heatmaps organized in columns. Each column represents a specific type of flow, while the vertical axis provides the identifying name of the corresponding region. The width of each column is proportional to the number of temporal observations. The ranking's position is represented by shades of red, with higher saturation indicating a higher ranking position.
This comprehensive visualization reveals a general trend of rank stability, particularly among top-ranked regions, which tend to maintain their positions across various flow types and years. However, we observe notable exceptions to this pattern, with certain regions demonstrating high centrality only in specific layers. Middle-ranked regions exhibit greater heterogeneity in their rankings across different flow types and years, indicating more dynamic centrality patterns in this tier.

Further analysis of PageRank trends (in the specific instance of the Migration layer) reveals that London maintains a strong upward trajectory in centrality through time, despite a temporary decline in 2016 (likely due to the Brexit referendum), highlighting the city's resilience as a key migration hub (see Supplementary Information for details).

To distill key information from these temporal trends, Tables \ref{tab:rankings_1} and \ref{tab:rankings_2} present a focused analysis of PageRank ranking dynamics. We highlight regions with the highest and lowest average rankings, as well as those experiencing the most significant increases and decreases in ranking positions across layers. This analysis uncovers that certain regions, such as Ile-de-France, consistently maintain high centrality across multiple flow types, demonstrating their multifaceted importance in European networks. Conversely, other regions, like Lazio, exhibit exceptional centrality in specific domains, suggesting specialized roles within particular flow networks.

\begin{table}[t!]
\centering
\footnotesize
\caption{Multiplex rankings (2010-2018): highest and lowest average, largest increases and decreases (excluding autonomous cities, Åland Islands, Atlantic island regions, and French overseas departments).}
\label{tab:multi_rankings}
\begin{tabular}{l@{\hspace{10pt}}r}
\toprule
\textbf{Category} & \textbf{Region} \\
\midrule
Highest Average: 1 & Ile-de-France (FR) \\
Highest Average: 2 & Comunidad de Madrid (ES) \\
Highest Average: 3 & Noord-Holland (NL) \\
Highest Average: 4 & Cataluña (ES) \\
Highest Average: 5 & Lombardia (IT) \\
\addlinespace
Lowest Average: 1 & Liechtenstein (LI) \\
Lowest Average: 2 & Molise (IT) \\
Lowest Average: 3 & La Rioja (ES) \\
Lowest Average: 4 & Flevoland (NL) \\
Lowest Average: 5 & Voreio Aigaio (EL) \\
\addlinespace
Largest Increase: 1 & Bratislavský kraj (SK) \\
Largest Increase: 2 & Leipzig (DE) \\
Largest Increase: 3 & Alentejo (PT) \\
Largest Increase: 4 & Kypros (CY) \\
Largest Increase: 5 & Nord-Vest (RO) \\
\addlinespace
Largest Decrease: 1 & Dytiki Ellada (EL) \\
Largest Decrease: 2 & Pohjois- ja Itä-Suomi (FI) \\
Largest Decrease: 3 & West Central Scotland (UK) \\
Largest Decrease: 4 & Northern Ireland (UK) \\
Largest Decrease: 5 & Länsi-Suomi (FI) \\
\bottomrule
\end{tabular}
\end{table}

Table \ref{tab:multi_rankings} presents an analysis of the multiplex PageRank ranking across regions. It highlights regions with the highest and lowest average rankings, as well as those experiencing the most substantial positive and negative shifts in their ranking positions. 
The analysis reveals that Ile-de-France, Comunidad de Madrid, Noord-Holland, Cataluña, and Lombardia consistently maintain the highest average rankings in the multiplex network. This suggests that these regions play central roles across multiple types of flows within the European network.
Conversely, we observe significant upward mobility in the rankings for regions such as Bratislava and Leipzig. These regions demonstrate the most substantial improvements in their multiplex PageRank positions, indicating an increase in their overall importance within the interconnected flow networks over time.

\subsubsection*{Single-Layer vs. Multiplex}
To quantify the insights offered by the multilayer approach, we compare the regional rankings derived from the multiplex network against a simpler, aggregate baseline. This baseline is created by averaging the rankings of the single-layer PageRank for each region. Such single-layer averaging sets a baseline and, at the same time, represents a composite index of importance---a region's overall status is treated as the sum of its parts across different domains.
The difference between a region's rank in the multilayer analysis and its rank in this simplified average baseline is therefore a powerful indicator of multiplex effects. It reveals how a region's ability to integrate different flow types enhances or diminishes its overall importance in a way that simple aggregation cannot capture.

\begin{figure}[t]
    \centering
    \includegraphics[width=0.7\textwidth]{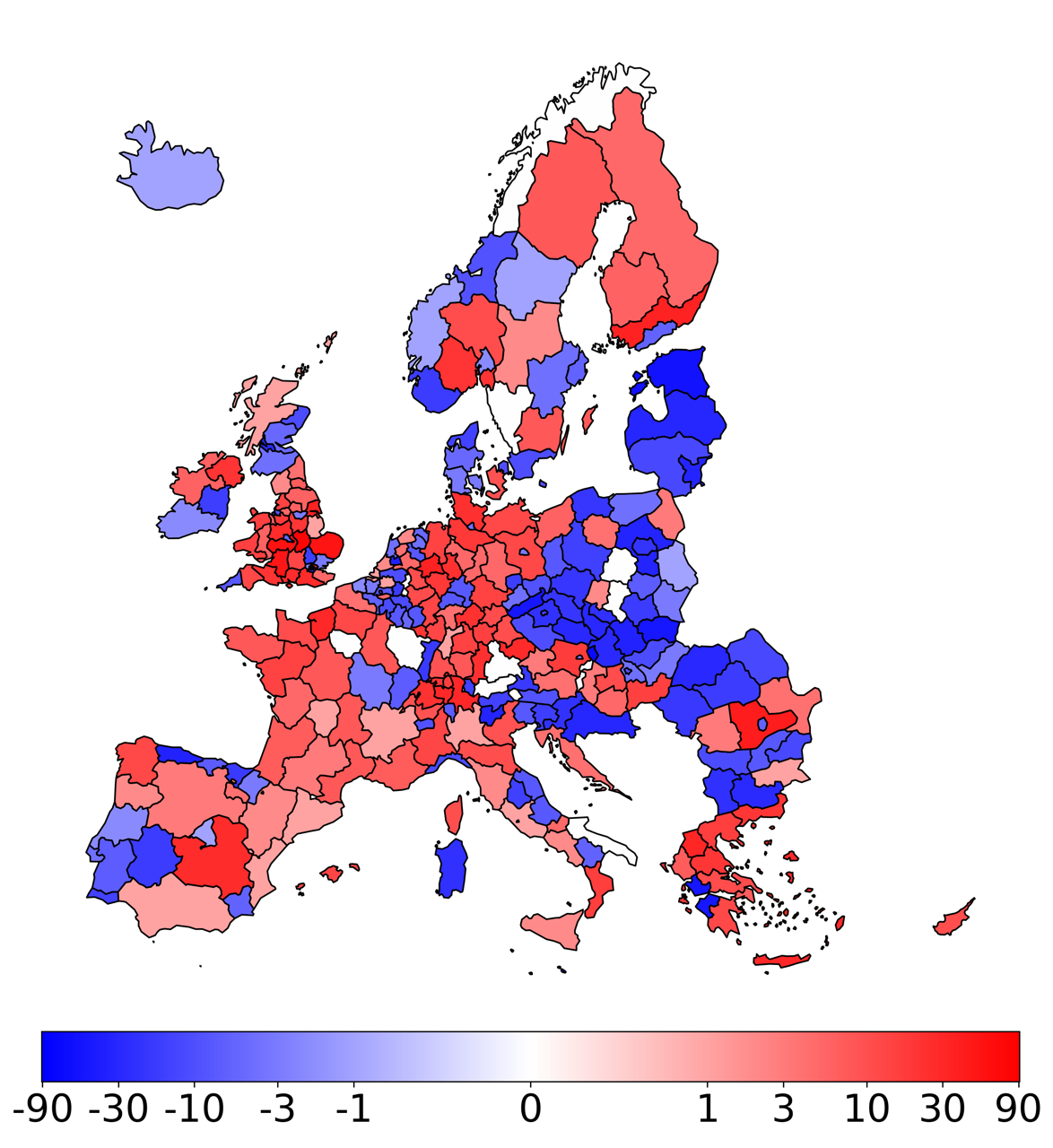}
    \caption{Change in node ranking: multilayer vs. average single-layer PageRank. Positive values (red) indicate an increase in ranking in the multilayer PageRank compared to the average of single-layer PageRanks, while negative values (blue) indicate a decrease. The intensity of the color represents the number of positions changed, with darker shades indicating larger changes. A logarithmic scale is used to emphasize changes near zero while still representing larger changes.}
    \label{fig:map_changes}
\end{figure}

Figure \ref{fig:map_changes} illustrates the changes in node ranking when comparing these two approaches. A comprehensive explanation is provided below.
The single-layer PageRank rankings were averaged to obtain a composite ranking, representing each node's average importance across all layers.
The positions of nodes in the multilayer PageRank ranking were compared to their positions in the average single-layer PageRank ranking. The difference in ranking positions was calculated for each node.
Nodes exhibiting an increase in ranking are indicated by positive values (red). This denotes an increase in the node's ranking in the multilayer PageRank relative to the average single-layer PageRank. That is, the node is more important in the multilayer analysis.
Negative values (blue) indicate a reduction in the node's ranking within the multilayer PageRank relative to the average single-layer PageRank. This suggests that the node is of less importance in the multilayer analysis.
This visualization reveals substantial variations, underscoring the importance of considering multiplex centrality measures to obtain comprehensive information not discernible from individual-layer analyses and revealing a heterogeneous pattern of ranking shifts. Notably, within individual countries, we observe both positive and negative shifts in regional rankings. The most consistent shift is observed for Malta, with a remarkable change of 91 positions, underscoring the potential for significant discrepancies between single-layer and multiplex centrality measures.

An analysis of the correlations between single-layer and multiplex PageRank values (see Supplementary Information) reveals moderate positive relationships.
While a perfect correlation would imply that the multilayer analysis is redundant, and no correlation would suggest that single-layer importance is irrelevant, the observed moderate correlation shows that while importance in a single, dominant flow contributes to a region's overall standing, it is not the sole determinant. The multilayer approach thus manages to capture the collective effect of diverse flow types, which is not fully visible from any single perspective.

\subsection{Community Detection: Infomap}

The application of the Infomap algorithm to our multiplex network revealed a complex community structure across European regions (Figure \ref{fig:communities}). 
This community detection algorithm identifies cohesive groups of regions based on the intensity of their mutual interactions across all flow types. A community represents a group of regions whose internal interactions are stronger than their connections to the rest of the network. These communities can be interpreted as empirically derived functional regions, i.e., economic and social subsystems whose boundaries are defined by flows, not by formal administrative lines.
A total of 82 communities were identified, which appears to be a reasonable number given that the total number of regions and countries involved is approximately 300 and 30, respectively. This equates to an average of approximately two and a half communities per country.
Our analysis reveals that these communities exhibit a mix of national cohesion and cross-border associations. This finding provides direct evidence for the dual nature of the European system. On the one hand, the enduring power of the nation-state remains a primary organizing force for many regional interactions. On the other hand, the emergence of strong cross-border communities signifies the development of transnational corridors of integration. The value of our approach lies in its ability to map the geography of this complex, multi-factor structure.
Notable cases that emerged include the following:

\begin{figure}[t]
    \centering
    \includegraphics[width=0.7\textwidth]{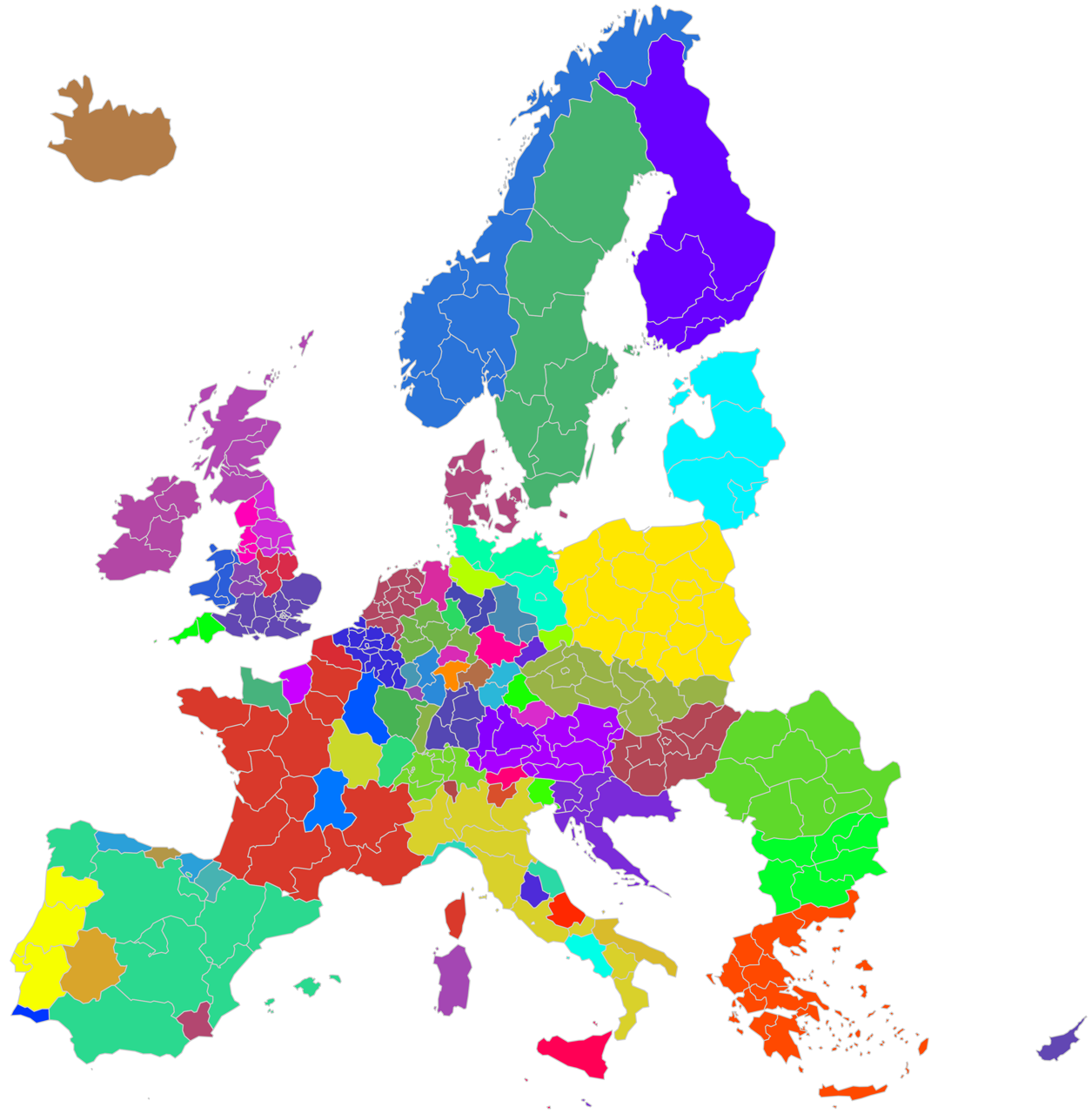}
    \caption{This map displays the NUTS2 regions of Europe, with the different colors representing the 82 distinct communities detected by the Infomap algorithm applied to the multiplex network for the year 2010.}
    \label{fig:communities}
\end{figure}

\begin{itemize}
    \item Belgium forms a community with Luxembourg and a neighboring Dutch region, suggesting strong economic and social ties in this cross-border area.
    \item The Czech Republic and Slovakia form a single community, reflecting their historical and ongoing close relations.
    \item A community comprises many English regions and Cyprus, indicating strong connections despite geographical distance.
    \item Several countries form predominantly self-contained communities, including Romania, Austria, Poland, Greece, Portugal, Hungary, Denmark, the Netherlands, Norway, Bulgaria, Finland, Malta, and Iceland. This suggests that these nations have stronger internal than external flows across the analyzed dimensions.
    \item Spain, France, and Italy each display a core community of multiple regions, with additional smaller communities, indicating complex internal structures.
    \item Cross-border communities are observed between Åland (Finland) and Sweden, as well as between Liechtenstein and Switzerland, highlighting strong regional ties that transcend national borders.
    \item The Baltic states (Lithuania, Estonia, and Latvia) form a cohesive community, reflecting their geographical proximity and shared historical background.
    \item Germany and the United Kingdom exhibit highly fragmented community structures, suggesting complex and diverse flow patterns within these countries.
    \item Slovenia, Croatia, and Malta form an unexpected community, potentially indicating strong economic or social ties among these Mediterranean and Adriatic regions.
    \item Northern Ireland and Ireland constitute a single community, aligning with their geographical proximity and historical connections.
\end{itemize}

Having 82 communities poses a set of challenges for their representation on a colored map. For more detailed information, we report the full listing of the communities in Table \ref{tab:comm_res} of the Supplementary Information.

\section{Discussion}\label{discussion}

Our analysis, grounded in a multilayer network framework, contributes to several key discussions in regional science and economic geography. By moving beyond single-flow analyses, we provide empirical weight to the theoretical arguments for a relational understanding of regional economies~\cite{amin2004regions,bathelt2011relational} and offer a more nuanced map of the \emph{space of flows}
within Europe~\cite{castells1996information}.

A primary finding is the empirical validation of a complex core--periphery structure in Europe's functional network. The heavy-tailed distributions we observe are a common signature of complex systems~\cite{albert2002statistical}, but our analysis gives this abstract property a concrete geographical meaning. The dominance of a few versatile hubs aligns with research on the world city network
, which finds that a limited number of cities act as primary nodes for global capital and knowledge~\cite{taylor2015world}. Our work shows how this hierarchy exists not just for global cities but also operates at the NUTS2 regional level across a more diverse set of~interactions.

The second key finding of our study is that moving from a single-layer to a multilayer centrality analysis fundamentally refines our understanding of regional importance. The primary contribution of the multiplex approach is its ability to look beyond a region’s performance in one domain and assess its overall systemic role, distinguishing between two types of regions---specialized and versatile.
While a single-layer analysis can clearly identify a region's specific strengths---for instance, the strong attractive power of German regions as specialized hubs for migration flows---it cannot, by itself, determine if this specialization translates into broader systemic influence. The multiplex perspective addresses this directly. We quantify this by measuring the rank shift, which is the change in a region's importance when moving from a simple average of its single-layer ranks to its integrated rank in the multilayer network. This shift is a powerful indicator of non-linear, synergistic effects, revealing how a region's ability to combine diverse flows of capital, knowledge, and people enhances or diminishes its overall importance.
This method yields significant empirical insights. The most pronounced ranking changes, for instance, are observed in Eastern Europe. 
This finding does not simply indicate growth, but rather points to a deep, structural transformation. The dramatic shifts suggest that the economic integration of these regions is not a shallow phenomenon, but a comprehensive evolution that is reshaping their role across the entire European system. This volatility means that as these regions integrate more deeply into the European system, they are actively sorting into distinct functional roles.

Ultimately, the practical value of this approach lies in its ability to reveal the functional advantages of versatility over specialization, thereby providing a more realistic map of Europe's economic geography. It demonstrates that regional importance is a multi-faceted characteristic, not a single, unified concept. The ability to effectively bridge different economic and social dimensions constitutes a distinct and crucial form of centrality that conventional analyses, which examine each flow independently, overlook. This offers a new lens to policymakers, presenting a strategic choice between development strategies that reinforce a successful specialization and those that foster the network versatility that appears to be a key feature of systemically important regions.

These general findings are further substantiated by specific case studies. Notably, Bratislava and Leipzig demonstrate the most significant increases in multiplex ranking from 2010 to 2018. This finding aligns with their documented economic trajectories during this period. Bratislava, despite a relatively stagnant population, experienced remarkable economic growth~\cite{HanzlWeiss2018BRATISLAVAAV}. Its GDP per capita at purchasing power parity surpassed that of Vienna, placing it among Europe's top 10 leading regions. This economic surge was primarily driven by substantial foreign direct investment, particularly in the automotive sector, leading to full employment in the region.
Similarly, Leipzig emerged as Germany's fastest-growing city in the 2010s~\cite{cudny2022growth}. Its remarkable growth can be attributed to massive public investments, subsidies, and support programs across various policy fields and sectors. These public initiatives were instrumental in mobilizing significant private capital investments across all urban sectors, fueling the city's rapid development.
These case studies of Bratislava and Leipzig illustrate how the multiplex analysis captures complex regional dynamics that might be missed in single-layer examinations. The multiplex approach effectively reflects the multifaceted nature of regional development, encompassing factors such as foreign investment, economic growth, and policy interventions, which collectively influence a region's centrality within the European network of flows.

Finally, our community detection results provide direct, empirical evidence on the long-standing debate between European integration and national cohesion, revealing a complex reality that is not an either/or scenario. On the one hand, the emergence of strong cross-border communities validates the concept of a Europe of regions, where functional economic units are defined by flows rather than formal borders. The integrated community spanning the island of Ireland or the cohesive block linking Belgium, Luxembourg, and a Dutch region are prime examples of this phenomenon, aligning with scholarly work on the rise in global city-regions~\cite{scott2001global}. On the other hand, our analysis simultaneously confirms the enduring power of the nation-state. The persistence of strong, self-contained national communities, even in highly globalized countries like Denmark and the Netherlands, demonstrates that the state remains the primary container for a dense web of socio-economic flows. Together, these findings depict a European space characterized by a dual structure---a mosaic of deeply integrated national systems overlaid with powerful, transnational corridors of interaction.
Beyond simply transcending administrative borders, our flow-based approach challenges an even more fundamental assumption---the primacy of geographic distance. By revealing non-intuitive communities, such as the link between English regions and Cyprus, the analysis demonstrates that intense relational proximity (e.g., through strong financial ties) can be far more significant for community formation than spatial proximity. 
More specifically, the England--Cyprus community result is no longer counter-intuitive when viewed through a non-geographic lens. The connection is a direct reflection of relational ties, most notably the strong, historic financial links between the UK and Cyprus, which is a major financial hub, and their shared status as Commonwealth members. The algorithm correctly identifies that the combined strength of these specific connections creates a more cohesive community than the ties between those English regions and many of their geographically closer European neighbors.
This finding offers a contrast to methodologies like spatial econometrics, which often presuppose that distance is the key determinant of interaction. By letting the data define the connections, our work uncovers the functional topology of Europe and provides an empirical method for analyzing the relational assets of regions, which is a central task for contemporary economic geography.
This growing understanding of relational proximity over distance not only reveals unexpected cross-regional connections but also sheds light on situations where historical and political divisions might suggest limited integration. Our results highlight dynamics that could remain hidden when using only traditional perspectives. For example, on the island of Ireland---an area marked in recent history by political conflict and border tensions---our flow-based analysis shows that Ireland and Northern Ireland function as a single integrated community. Although a purely historical or political lens might emphasize division, the empirical patterns in cross-border flows reveal a deep economic and social interdependence on the island. This means that even where legacy narratives predict separation, everyday connections can point towards unexpected forms of integration. Thus, our approach refines our understanding of territorial cohesion and serves as a tool to uncover the latent patterns of interaction throughout Europe.

In conclusion, this study demonstrates that a multilayer network approach offers a more nuanced and structurally aware perspective on European regional dynamics than traditional single-layer analyses. Our primary contribution is the ability to move beyond simple measures of importance and empirically map Europe's complex functional geography. We have shown that this method can distinguish between specialized regions, dominant in a single domain, and versatile ones, whose importance is derived from their ability to effectively bridge multiple types of flows. This distinction, exemplified by the significant rise in the importance of regions like Bratislava and Leipzig in the multiplex analysis, reveals a new dimension of regional strength rooted in systemic integration rather than specialization in one sector.
However, we must acknowledge the limitations of this study, which primarily stem from the nature of the available data. Our analysis is constrained by the quality and consistency of the ESPON dataset. The harmonization of diverse, cross-national flow data is an immense challenge, and potential inaccuracies or biases in the original data collection could influence the results. Furthermore, while our network analysis is powerful at revealing unexpected structural patterns, it identifies correlation, not causation. These findings are therefore best understood as a data-driven map that generates new, specific hypotheses for future qualitative and econometric investigation, which would be needed to explain the causal mechanisms driving the observed connections.
Despite these limitations, our findings carry significant policy implications. First, the distinction between versatility and specialization can serve as a powerful tool for tailoring regional development strategies. Rather than simply identifying a single best type of region to support, this analysis allows policymakers to design more nuanced interventions; for highly specialized regions, policy could focus on mitigating risks by fostering diversification, while for versatile regions, the focus could be on leveraging their bridging role to benefit the wider network. Second, the identification of robust cross-border communities suggests that cohesion policy could be more effective if it were partially targeted towards these empirically defined functional regions, rather than being restricted by purely administrative boundaries. Fostering cooperation along these true corridors of interaction could accelerate integration and provide a greater return on investment.
Ultimately, this work provides both a method and a new lens for policymakers to better understand and navigate the complex, interconnected reality of the European space. Future research should move towards explaining the structures we have identified. This could involve qualitative case studies to unpack the specific economic and historical drivers behind non-intuitive communities, or the use of temporal network models to investigate how specific policy interventions or external shocks propagate through this multilayer system and reshape regional connectivity over time.
Finally, it is worth considering how these network structures might relate to broader socio-economic phenomena, such as immigration from extra-EU countries. Regions with high multilayer centrality are those that, in the observation period (and before), attracted and integrated a significant non-EU population. Among these top-ranked hubs, we find Île-de-France, Comunidad de Madrid, Cataluña, North Holland, and Lombardy, recognised as primary gateways for extra-EU migration ({\url{https://www.immigration.interieur.gouv.fr/Info-ressources/Etudes-et-statistiques/Les-chiffres-de-l-immigration-en-France/Population-immigree-par-departement?} accessed on 16/09/2025; \url{https://www.ine.es/dyngs/Prensa/es/ECP1T24.htm} accessed on 16/09/2025; \url{https://www.istat.it/wp-content/uploads/2024/10/REPORT-CITTADINI-NON-COMUNITARI_Anno-2023.pdf?} accessed on 16/09/2025; \url{https://www.cbs.nl/en-gb/dossier/asylum-migration-and-integration/how-many-residents-of-the-netherlands-have-a-non-dutch-background-?} accessed on 16/09/2025}). As a future development of this paper, this suggests the consideration of a multiplier effect, whereby the immigrant workforce, often concentrated in essential low- to medium-skilled sectors, provides a ``system elasticity'' that eases labor shortages and creates a robust foundation upon which high-productivity sectors can thrive. In turn, this might boost the very capital, knowledge, and tourism flows that define a region's systemic importance. Conversely, the North Aegean region in Greece, which experienced high numbers of non-EU arrivals due to humanitarian crises ({\url{https://www.migrationpolicy.org/article/refugee-flows-lesvos-evolution-humanitarian-response} accessed on 16/09/2025}), does not rank high according to our analysis. We argue that the immigrant population is not integrated into the socio-economic system of the region. This distinction underscores that our findings appear to be related to the economic integration of a stable immigrant population---a stock---rather than to transitory flows, highlighting another dimension of functional connectivity within the European space.

\begin{credits}
\subsubsection{Author Contributions.} Conceptualization: E.C. and A.F.; methodology: E.C. and A.F.; software: E.C.; validation: E.C.; formal analysis: E.C.; investigation: E.C.; resources: A.F.; data curation: E.C.; writing---original draft preparation: E.C.; writing---review and editing: E.C. and A.F.; visualization: E.C. and A.F.; supervision: A.F.; project administration: A.F.; funding acquisition: A.F. All authors have read and agreed to the published version of the manuscript.

\subsubsection{Funding.} A.F. is supported by SoBigData.it, which receives funding from European Union – NextGenerationEU – National Recovery and Resilience Plan (Piano Nazionale di Ripresa e Resilienza, PNRR) – Project: “SoBigData.it – Strengthening the Italian RI for Social Mining and Big Data Analytics” – Prot. IR0000013 – Avviso n. 3264 del 28/12/2021.

\subsubsection{Data Availability Statement.} All data used are publicly available \cite{EsponDatabase}. Moreover, data and code are available in the Zenodo repository \cite{Calo_2025}.

\subsubsection{\ackname}
We thank the IMT Lucca Networks research unit, Leonardo Ialongo, Simone Daniotti, Davide Fiaschi, Angela Parenti, and Cristiano Ricci for providing insightful discussions. We acknowledge the use of LLMs for text refining. The authors reviewed and edited the content as needed and take full responsibility for the content of the publication.
 
\subsubsection{\discintname}
The authors declare no conflicts of interest.
\end{credits}

\begin{table}[t]
\caption{Overview of flow type, data sources, methodologies, and our analysis.}
\label{tab:flow_overview}
\begin{adjustbox}{center}
\small
\begin{tabular}{|p{2cm}|p{2.5cm}|p{2cm}|p{4.5cm}|p{2.5cm}|}
\hline
\textbf{Flow Type} & \textbf{Description} & \textbf{Sources} & \textbf{Methodology} & \textbf{Our Analysis} \\
\hline
\textbf{Migration} & Number of people migrating between NUTS 2 regions & EUROSTAT and NSI & Multi-step process: Base Data, Stock Gain estimation, In-Out-Cross estimation. Country-to-country matrices created, gaps filled using stock-gain method and linear models. Region-to-region flows estimated by decomposing country-level data& Applied floor function to ensure whole numbers. Divided by origin region population  (see SI)\\
\hline
\textbf{Tourism} & Number of tourists traveling between NUTS 2 regions & EUROSTAT and UNWTO for country-to-country; EUROSTAT for regional domestic arrivals & Completed gaps in country-to-country and disaggregated country-to-country to region-to-region. Methods include cross-referencing UNWTO indexes, interpolation/extrapolation, and gravity model analysis using GDP, arrivals, and distance data& Applied floor function. Divided by origin region population\\
\hline
\textbf{FDI} & Shareholders' funds (thousand euros) in foreign-owned companies & AMADEUS database (Bureau van Dijk) & Aggregated firm-level data. Included intraregional and interregional intra-national flows& Summed FDI across all sectors. Divided by origin region GDP (see SI)\\
\hline
\textbf{Remittances} & Regionalized bilateral remittance estimates (thousand euros)& EUROSTAT and World Bank & Estimated regional-level flows by regionalizing national-level data using ratio of regional to national migration flows& Divided by origin region GDP\\
\hline
\textbf{Freight Transport} & Total freight flow between NUTS-2 regions (thousand tons) & Various, for road, rail, maritime, and air transport & Performed consistency and plausibility checks. Developed disaggregation procedures where regionalized data unavailable& Divided by total outgoing flows from each region, multiplied by region's relative economic importance\\
\hline
\textbf{Erasmus Student Mobility} & Higher education student mobility between partner countries & European Commission datasets & Geocoded individual movements to NUTS-2 regions based on sending and receiving institutions& Divided by origin region population\\
\hline
\textbf{Horizon 2020 Partnerships} & Number of H2020 partnerships between NUTS-2 regions & CORDIS project and participant organization lists & Geocoded organizations to NUTS-2 regions. Counted partnerships with coordinating partners as senders and other partners as receivers. & No additional processing\\
\hline
\textbf{Passenger Transport} & Passenger flows between NUTS-2 regions for air, maritime, and rail transport & Various Eurostat datasets & Implemented appropriate disaggregation procedures where regionalized data unavailable& Summed air, maritime, and rail (×1000) passenger flows. Divided by origin region population\\
\hline
\end{tabular}
\end{adjustbox}
\end{table}

\begin{sidewaystable}
\centering
\caption{Single-layer rankings (2010-2018): highest and lowest average, largest increases and decreases for Erasmus, FDI, Freight, and Horizon2020 (excluding Melilla and Ceuta autonomous cities; Açores, Madeira, Åland, Canarias and Baleares island regions, and French overseas departments). E=East, N=North. Note: For FDI lowest average, other Danish regions, Inner London - East, and several Greek regions are omitted due to equal last position (zero inflows). Four Norwegian regions (Oslo og Akershus, Nord-Norge, Vestlandet, Trøndelag) are absent from the FDI network. For Erasmus largest increase and lowest average, regions absent from the network in some years are omitted from the rankings.}\label{tab:rankings_1}
\begin{tabular}{lrrrr}
\toprule
Category & Erasmus & FDI & Freight & Horizon2020 \\
\midrule
Highest Average: 1 & Comunidad de Madrid (ES) & Noord-Holland (NL) & Lombardia (IT) & Ile-de-France (FR) \\
Highest Average: 2 & Ile-de-France (FR) & Ile-de-France (FR) & Zuid-Holland (NL) & Comunidad de Madrid (ES) \\
Highest Average: 3 & Andalucía (ES) & Luxembourg (LU) & Cataluña (ES) & Oberbayern (DE) \\
Highest Average: 4 & Cataluña (ES) & Eastern and Midland (IE) & Ile-de-France (FR) & Région de Bruxelles (BE) \\
Highest Average: 5 & Berlin (DE) & Comunidad de Madrid (ES) & Nord-Pas de Calais (FR) & Lazio (IT) \\
\addlinespace
Lowest Average: 1 & Sterea Ellada (EL)	 & Hovedstaden (DK) & Liechtenstein (LI) & Liechtenstein (LI) \\
Lowest Average: 2 & Yugoiztochen (BG) & Thessalia (EL) & Ísland (IS) & Valle d'Aosta (IT) \\
Lowest Average: 3 & Dytiki Makedonia (EL) & Espace Mittelland (CH) & Malta (MT) & Warmińsko-mazurskie (PL) \\
Lowest Average: 4 & Valle d'Aosta (IT) & Ticino (CH) & Sostinės regionas (LT) & Cumbria (UK) \\
Lowest Average: 5 & Warmińsko-mazurskie (PL) & Molise (IT) & Highlands and Islands (UK) & Opolskie (PL) \\
\addlinespace 	
Largest Increase: 1 & Kontinentalna Hrvatska (HR) & Bretagne (FR) & Kontinentalna Hrvatska (HR) & Extremadura (ES) \\
Largest Increase: 2 & Jadranska Hrvatska (HR) & East Yorkshire (UK) & Ionia Nisia (EL) & Leipzig (DE) \\
Largest Increase: 3 & Wielkopolskie (PL) & Zachodniopomorskie (PL) & Střední Morava (CZ) & Zentralschweiz (CH) \\
Largest Increase: 4 & Podlaskie (PL) & Midi-Pyrénées (FR) & Jihovýchod (CZ) & Prov. West-Vlaanderen (BE) \\
Largest Increase: 5 & Pomorskie (PL) & Yugoiztochen (BG) & Östra Mellansverige (SE) & Nord-Vest (RO) \\
\addlinespace
Largest Decrease: 1 & Nordjylland (DK) & Región de Murcia (ES) & N Ireland (UK) & Dél-Alföld (HU) \\
Largest Decrease: 2 & Inner London E (UK) & Sardegna (IT) & Nyugat-Dunántúl (HU) & Basilicata (IT) \\
Largest Decrease: 3 & Trentino (IT) & Dytiki Ellada (EL) & Peloponnisos (EL) & Moravskoslezsko (CZ) \\
Largest Decrease: 4 & Sjælland (DK) & Lancashire (UK) & Dél-Alföld (HU) & Malta (MT) \\
Largest Decrease: 5 & Liguria (IT) & Brandenburg (DE) & Etelä-Suomi (FI) & Jihovýchod (CZ) \\
\bottomrule 			
\end{tabular}
\end{sidewaystable}

\begin{sidewaystable}
\centering
\caption{Single-layer rankings (2010-2018): highest and lowest average, largest increases and decreases for Migration, Passengers, Remittances, and Tourism (excluding Melilla and Ceuta autonomous cities; Açores, Madeira, Åland, Canarias and Baleares island regions, and French overseas departments). E=East, W=West, NW=North West, NE=North East.}\label{tab:rankings_2}
\begin{tabular}{lrrrr}
\toprule
Category & Migration & Passengers & Remittances & Tourism \\
\midrule
Highest Average: 1 & Oberbayern (DE) & Ile-de-France (FR) & Ile-de-France (FR) & Jadranska Hrvatska (HR) \\
Highest Average: 2 & Stuttgart (DE) & Inner London W (UK) & Cataluña (ES) & Cataluña (ES) \\
Highest Average: 3 & Inner London E (UK) & Comunidad de Madrid (ES) & Luxembourg (LU) & Ile-de-France (FR) \\
Highest Average: 4 & Darmstadt (DE) & Inner London E (UK) & Comunidad de Madrid (ES) & Andalucía (ES) \\
Highest Average: 5 & Ile-de-France (FR) & Oberbayern (DE) & Rhône-Alpes (FR) & Rhône-Alpes (FR) \\
\addlinespace
Lowest Average: 1 & Liechtenstein (LI) & Liechtenstein (LI) & Liechtenstein (LI) & Liechtenstein (LI) \\
Lowest Average: 2 & Valle d'Aosta (IT) & Ipeiros (EL) & Voreio Aigaio (EL) & Prov. Brabant Wallon (BE) \\
Lowest Average: 3 & Molise (IT) & Burgenland (AT) & Ionia Nisia (EL) & Outer London W-NW (UK)\\
Lowest Average: 4 & Malta (MT) & Corse (FR) & Highlands and Islands (UK) &  Molise (IT)\\
Lowest Average: 5 & Basilicata (IT) & Dytiki Makedonia (EL) & NE Scotland (UK) & Outer London E-NE (UK)\\
\addlinespace 
Largest Increase: 1 & Vidurio ir vakarų Lietuvos (LT) & Etelä-Suomi (FI) & Ísland (IS) & Alentejo (PT) \\
Largest Increase: 2 & Sud-Vest Oltenia (RO) & Brandenburg (DE) & Nyugat-Dunántúl (HU) & Zuid-Holland (NL) \\
Largest Increase: 3 & Nord-Vest (RO) & Notio Aigaio (EL) & Pest (HU) & Jihovýchod (CZ) \\
Largest Increase: 4 & Sud-Est (RO) & Leicestershire (UK) & Västsverige (SE) &  Overijssel (NL) \\
Largest Increase: 5 & Centru (RO) & Prov. Hainaut (BE) & Östra Mellansverige (SE) & Ísland (IS) \\
\addlinespace 
Largest Decrease: 1 & Zentralschweiz (CH) & Yugozapaden (BG) & Attiki (EL) & Sud-Est (RO) \\
Largest Decrease: 2 & Sicilia (IT) & Kontinentalna Hrvatska (HR) & Kentriki Makedonia (EL) & Lubuskie (PL) \\
Largest Decrease: 3 & Campania (IT) & Yugoiztochen (BG) & Zuid-Holland (NL) & Dorset and Somerset (UK) \\
Largest Decrease: 4 & Ostschweiz (CH) & Severoiztochen (BG) & Noord-Holland (NL) & Champagne-Ardenne (FR) \\
Largest Decrease: 5 & Calabria (IT) & Latvija (LV) & Jadranska Hrvatska (HR) & Bourgogne (FR) \\
\bottomrule
\end{tabular}
\end{sidewaystable}


\clearpage

\setcounter{section}{0}
\renewcommand{\thesection}{S\arabic{section}}
\renewcommand{\theHsection}{S\arabic{section}}

\chapter*{Supplementary Information}
\section{Data}
Detailed information on the data sources and processing methodologies used for harmonization and estimation of missing values can be found in \cite{IRiE}.

\subsection{Population}\label{pop}
Total population on January 1st, sourced from \cite{EsponDatabase}. Manual additions were made for Mariotte (France) and Ireland for specific years. 

We normalize the migration flows by dividing them by the population of the origin region to address a potential bias. According to the gravity model in migration studies, regions with larger populations tend to have higher absolute numbers of incoming and outgoing migrants. By normalizing the flows, we eliminate this population size bias, allowing for a more accurate comparison of regions' relative contributions to migration patterns. This approach is also applied to other types of flows in our study, ensuring consistent analysis across different flow categories.

\subsection{GDP}\label{gdp}
Gross Domestic Product at current market prices (million euros), sourced from \cite{EsponDatabase}. 2011 values were used for regions lacking 2010 data. Manual additions were made for Iceland and Liechtenstein.

\section{Methods}
We configure the Infomap algorithm with the following parameters:
\begin{itemize}
    \item \texttt{two\_level=False}: Clusters the optimal number of nested modules, accommodating both country-level clusters and single-region clusters.
    \item \texttt{num\_trials=100}: Number of outer-most loops to run before selecting the best solution.
    \item \texttt{flow\_model='rawdir'}: Determines node visitation rates based on the given direction and weight of edges, without using a PageRank algorithm.
    \item \texttt{entropy\_corrected=True}: Corrects for negative entropy bias in small samples (many modules).
    \item \texttt{multilayer\_relax\_rate=0.15}: Probability of relaxing the constraint to move only within the current layer (default value).
\end{itemize}
We applied the Infomap algorithm to the multiplex network structure to identify communities. In this analysis, it is possible for a region to be assigned to multiple communities across different layers. To resolve such cases and provide a definitive community assignment, we employed a frequency-based approach. Specifically, each region was ultimately assigned to the community in which it appeared most frequently across all layers.

\section{Results}
\subsection{Network properties}
\begin{table}[b]
\normalsize
\centering
\begin{tabular}{ll@{\hspace{10pt}}r@{\hspace{10pt}}r@{\hspace{10pt}}r}
\hline
Layer & Year & Nodes & Edges & Density (\%) \\
\hline
Migration & 2010 & 297 & 70797 & 80.53 \\
Migration & 2011 & 297 & 71535 & 81.37 \\
Migration & 2012 & 297 & 72459 & 82.42 \\
Migration & 2013 & 297 & 74502 & 84.75 \\
Migration & 2014 & 297 & 75342 & 85.70 \\
Migration & 2015 & 297 & 75580 & 85.97 \\
Migration & 2016 & 297 & 75955 & 86.40 \\
Migration & 2017 & 297 & 76044 & 86.50 \\
Migration & 2018 & 297 & 76281 & 86.77 \\
Tourism & 2010 & 297 & 87635 & 99.68 \\
Tourism & 2011 & 297 & 87649 & 99.70 \\
Tourism & 2012 & 297 & 87652 & 99.70 \\
Tourism & 2013 & 297 & 87665 & 99.72 \\
Tourism & 2014 & 297 & 87662 & 99.72 \\
Tourism & 2015 & 297 & 87669 & 99.72 \\
Tourism & 2016 & 297 & 87686 & 99.74 \\
Tourism & 2017 & 297 & 87682 & 99.74 \\
Tourism & 2018 & 297 & 87692 & 99.75 \\
Freight & 2010 & 297 & 46695 & 53.12 \\
Freight & 2011 & 297 & 47315 & 53.82 \\
Freight & 2012 & 297 & 46820 & 53.26 \\
Freight & 2013 & 297 & 46663 & 53.08 \\
Freight & 2014 & 297 & 46703 & 53.12 \\
Freight & 2015 & 297 & 44742 & 50.89 \\
Freight & 2016 & 297 & 43628 & 49.63 \\
Freight & 2017 & 297 & 43326 & 49.28 \\
Freight & 2018 & 297 & 42587 & 48.44 \\
Erasmus & 2010 & 265 & 21551 & 30.80 \\
Erasmus & 2011 & 264 & 22094 & 31.82 \\
Erasmus & 2012 & 270 & 23405 & 32.22 \\
Erasmus & 2013 & 272 & 24027 & 32.60 \\
Erasmus & 2014 & 274 & 24576 & 32.85 \\
\hline
\end{tabular}
\caption{Network statistics for Migration, Tourism, Freight, and Erasmus.}
\label{tab:network_stats_1}
\end{table}
\begin{table}[t]
\normalsize
\centering
\begin{tabular}{ll@{\hspace{10pt}}r@{\hspace{10pt}}r@{\hspace{10pt}}r}
\hline
Layer & Year & Nodes & Edges & Density (\%) \\
\hline
FDI & 2010 & 292 & 24872 & 29.27 \\
FDI & 2011 & 292 & 25028 & 29.45 \\
FDI & 2012 & 292 & 25178 & 29.63 \\
FDI & 2013 & 292 & 25309 & 29.79 \\
FDI & 2014 & 292 & 25360 & 29.85 \\
FDI & 2015 & 292 & 25422 & 29.92 \\
FDI & 2016 & 292 & 25483 & 29.99 \\
FDI & 2017 & 292 & 25521 & 30.03 \\
FDI & 2018 & 292 & 25441 & 29.94 \\
Passengers & 2010 & 297 & 12144 & 13.81 \\
Passengers & 2011 & 297 & 12486 & 14.20 \\
Passengers & 2012 & 297 & 12547 & 14.27 \\
Passengers & 2013 & 297 & 12604 & 14.34 \\
Passengers & 2014 & 297 & 12649 & 14.39 \\
Passengers & 2015 & 297 & 12588 & 14.32 \\
Passengers & 2016 & 297 & 12716 & 14.46 \\
Passengers & 2017 & 297 & 12849 & 14.62 \\
Passengers & 2018 & 297 & 12998 & 14.79 \\
Remittances & 2010 & 297 & 81375 & 92.56 \\
Remittances & 2011 & 297 & 81355 & 92.54 \\
Remittances & 2012 & 297 & 81410 & 92.60 \\
Remittances & 2013 & 297 & 81243 & 92.41 \\
Remittances & 2014 & 297 & 81172 & 92.33 \\
Remittances & 2015 & 297 & 81016 & 92.16 \\
Remittances & 2016 & 297 & 81236 & 92.41 \\
Remittances & 2017 & 297 & 81250 & 92.42 \\
Remittances & 2018 & 297 & 81250 & 92.42 \\
Horizon2020 & 2015 & 287 & 7482 & 9.12 \\
Horizon2020 & 2016 & 284 & 7135 & 8.88 \\
Horizon2020 & 2017 & 288 & 6899 & 8.35 \\
Horizon2020 & 2018 & 279 & 6475 & 8.35 \\
\hline
\end{tabular}
\caption{Network statistics for FDI, Passengers, Remittances, and Horizon2020.}
\label{tab:network_stats_2}
\end{table}

Table \ref{tab:network_stats_1} and Table \ref{tab:network_stats_2} present the network statistics across European NUTS-2 regions from 2010 to 2018. For each flow type, we report the number of nodes, which indicates the participating NUTS-2 regions in the network, as well as the number of edges, representing the connections between these regions where a connection signifies a non-zero flow. Additionally, we calculate the density of each network as the ratio of actual connections to the total possible connections, providing insight into how interconnected the regions are within each flow type.

\begin{figure}[t]
    \centering
    \includegraphics[width=\textwidth]{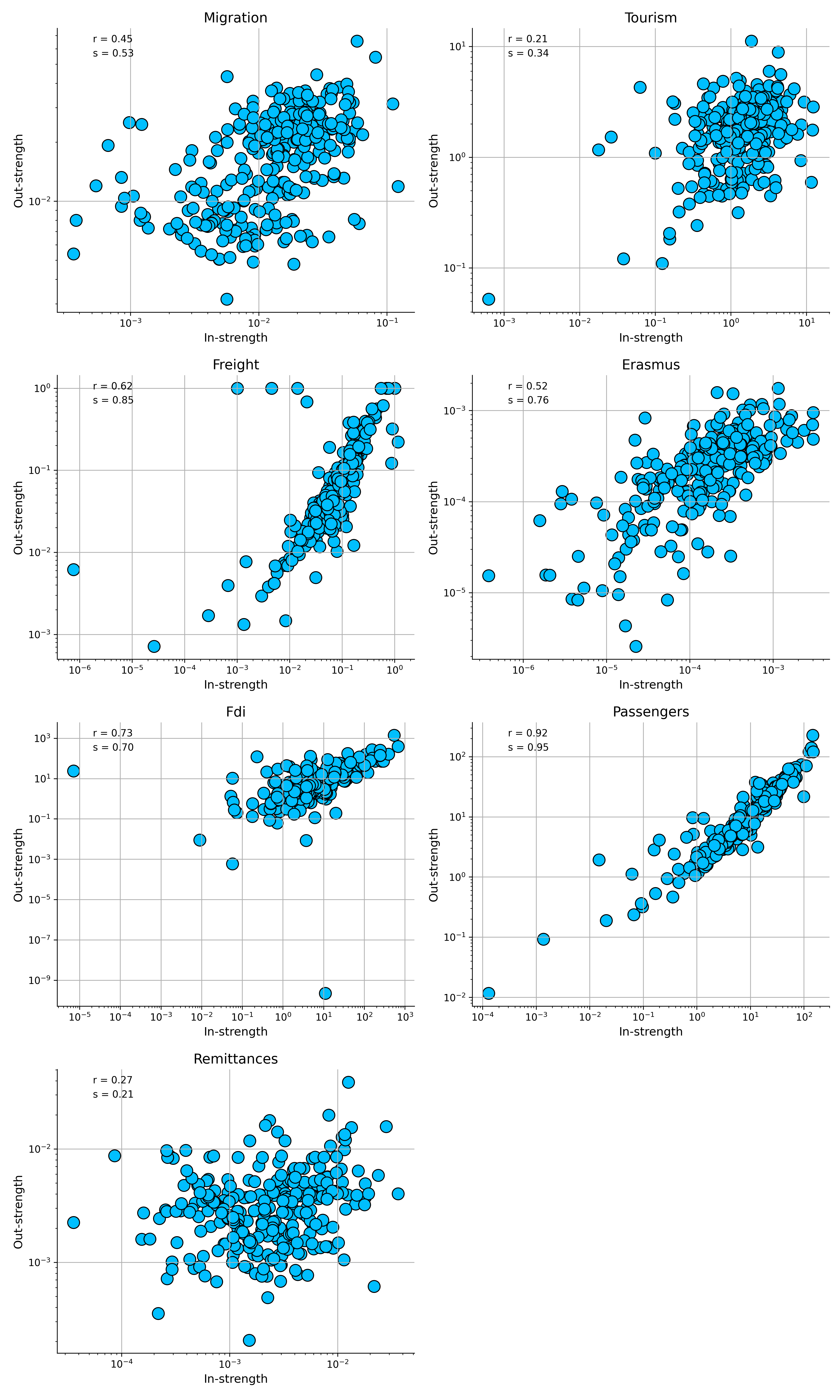}
    \caption{In-strength VS out-strength for the year 2010.}
        \label{fig:in_out_2010}
\end{figure}
\begin{figure}[t]
    \centering
    \includegraphics[width=\textwidth]{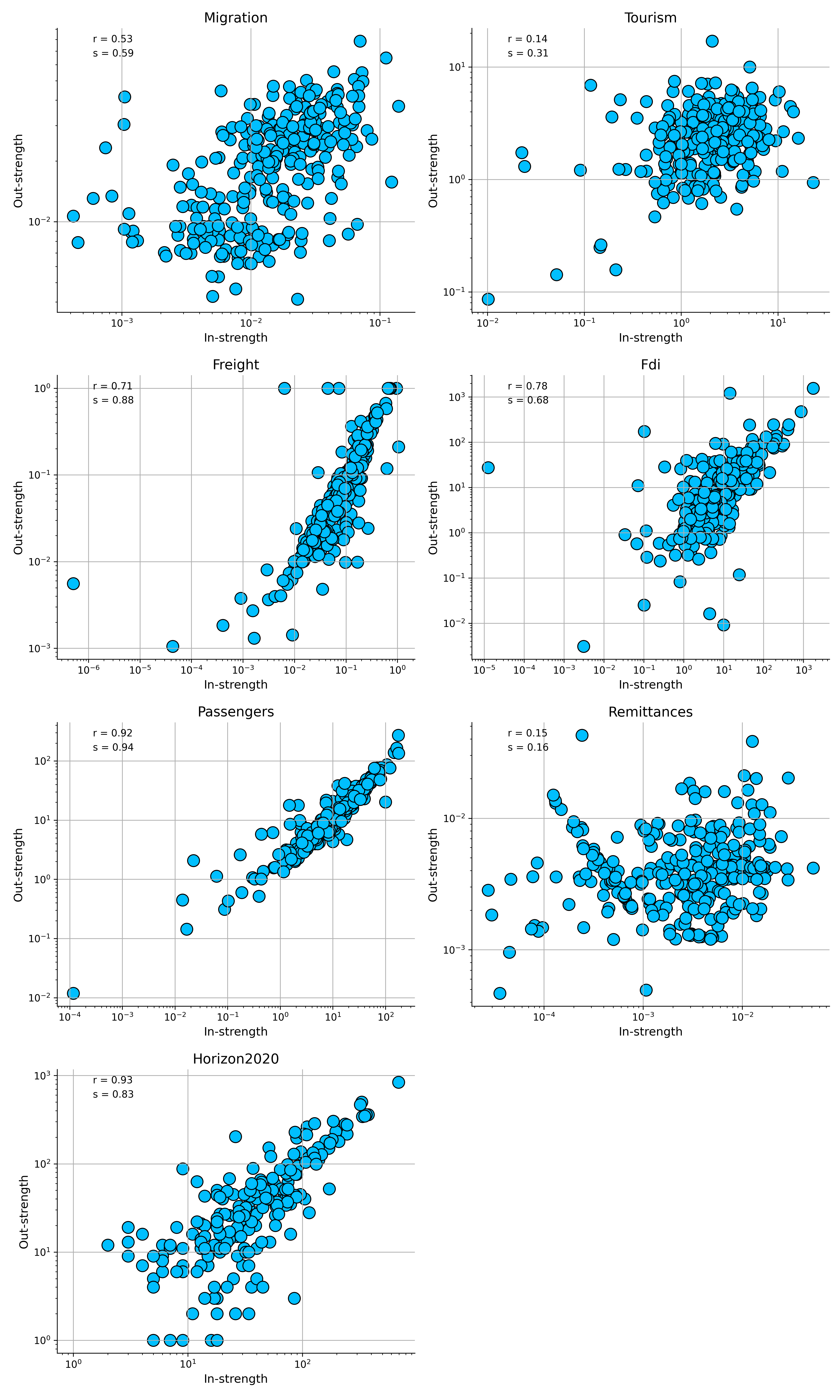}
    \caption{In-strength VS out-strength for the year 2018.}
        \label{fig:in_out_2018}
\end{figure}

Fig. \ref{fig:in_out_2010} and \ref{fig:in_out_2018} depict the relationship between in-strength and out-strength for all flow types in 2010 and 2018, respectively. These scatter plots reveal notable differences in correlations between various flow types, which remain consistent across both years. For instance, in 2010, the Spearman correlation coefficients range widely from 0.16 to 0.94, indicating diverse patterns of association between inflows and outflows across different domains. This substantial variation in correlations persists in 2018, suggesting that the underlying structures of these regional flow networks maintain their distinct characteristics over time.

\begin{figure}[t!]
    \centering
    \includegraphics[width=\textwidth]{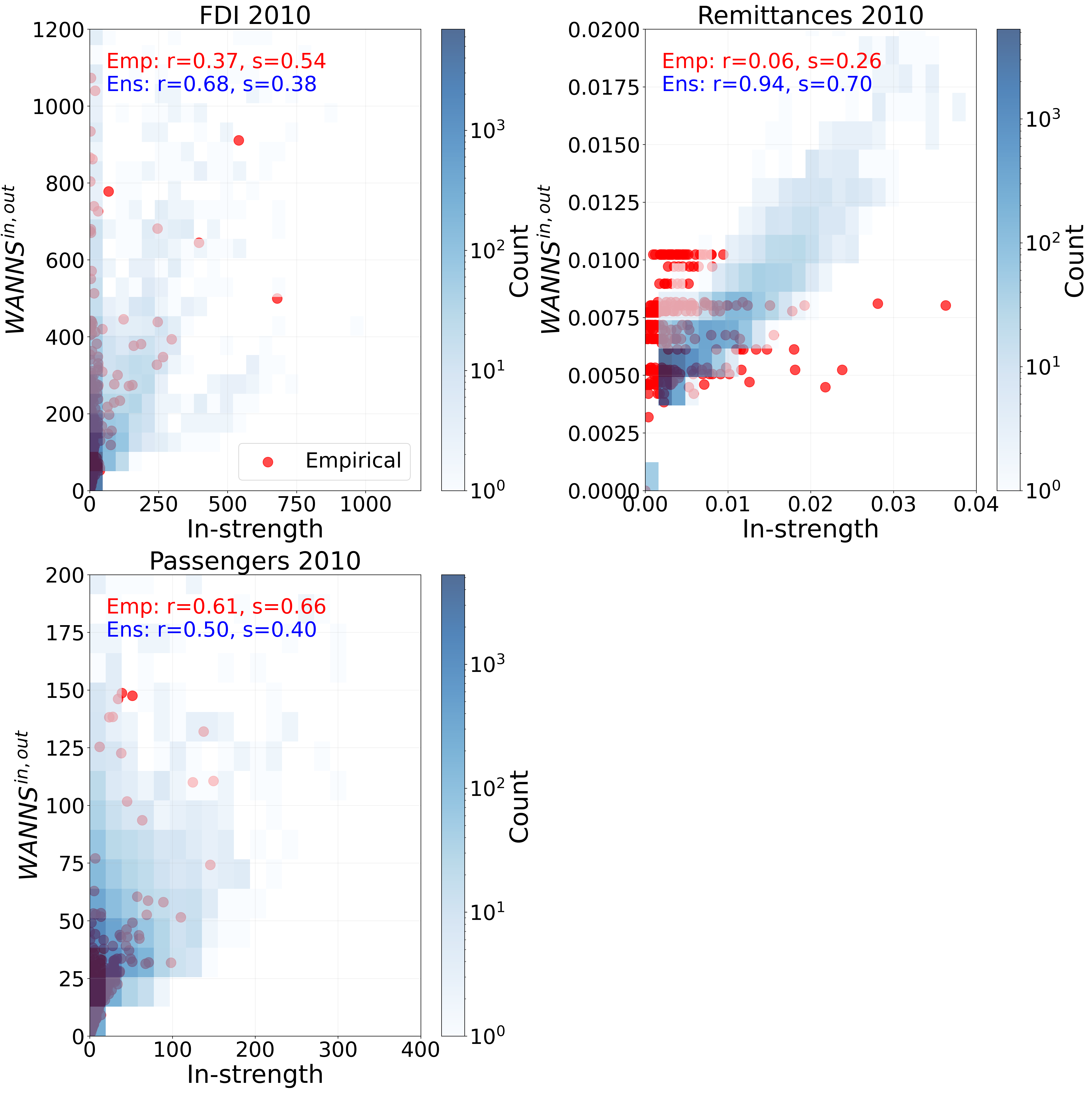}
    \caption{Comparison of empirical and ensemble WANNS for FDI, Remittances and Passengers in 2010. Each subplot shows the relationship between in-strength and $\textup{WANNS}^{in,out}$ values. The red points indicate empirical data, while the blue histogram represents the distribution of ensemble results. Correlation coefficients (Pearson's r and Spearman's s) are displayed for both empirical and ensemble data.}
    \label{fig:wanns_3}
\end{figure}

Fig. \ref{fig:wanns_3} presents the $\textup{WANNS}^{in,out}$ for the empirical networks alongside 50 realizations drawn from the null model ensembles for FDI, Remittances, and Passengers in 2010. 

\begin{figure}[t]
    \centering
    \includegraphics[width=\textwidth]{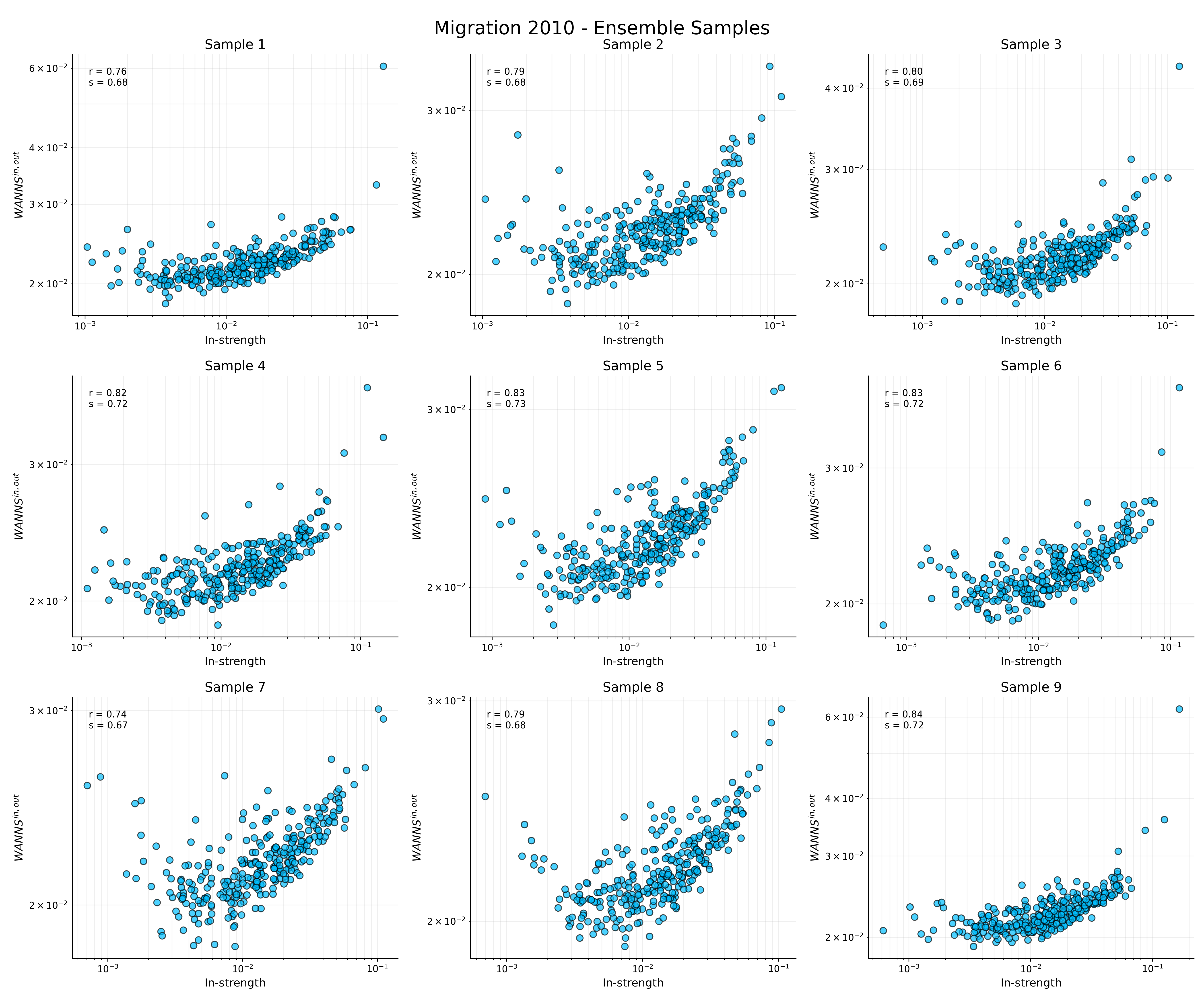}
    \caption{$\textup{WANNS}^{in,out}$ VS in-strength for ensemble copies for Migration the year 2010.}
        \label{fig:wanns_ens}
\end{figure}

Fig. \ref{fig:wanns_ens} focuses specifically on the Migration flow type in 2010, showing the relationship between the $\textup{WANNS}^{in,out}$ and the in-strength for ensemble copies. This visualization helps to understand the assortativity patterns in the migration network, revealing how regions with higher in-strength tend to receive flows from regions with higher out-strength.

\begin{figure}[t]
    \centering
    \includegraphics[width=\textwidth]{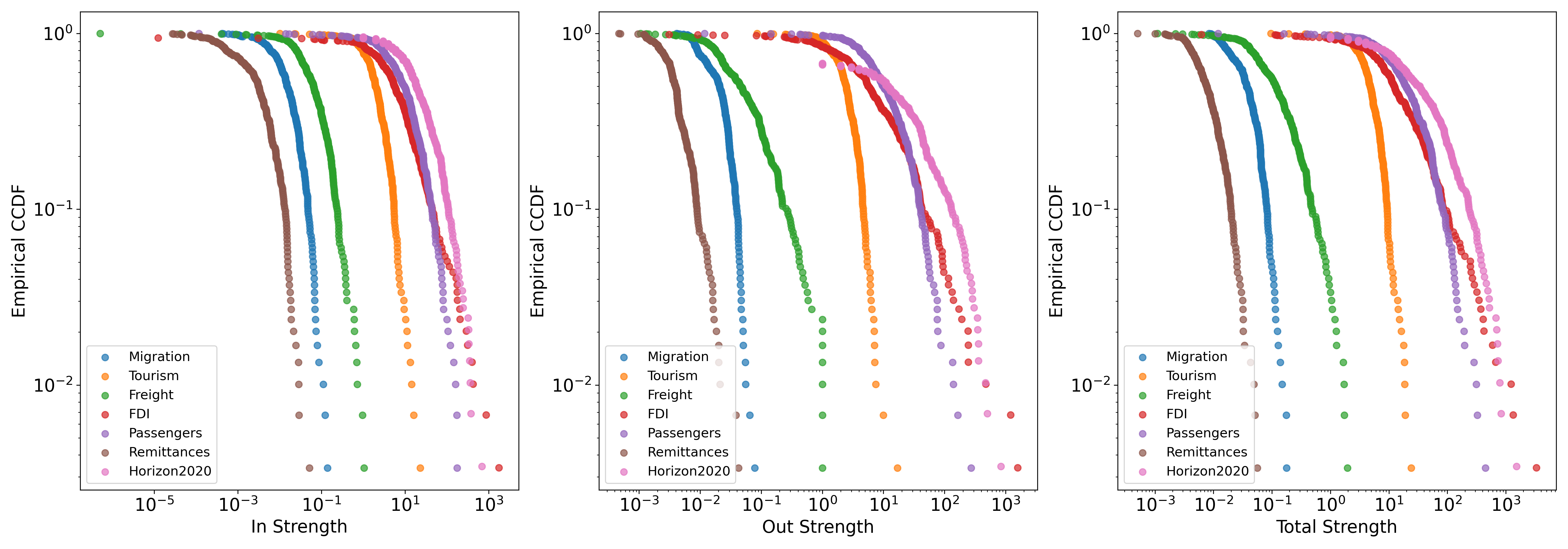}
    \caption{Complementary cumulative distribution function for the year 2018.}
        \label{fig:ccdf_2018}
\end{figure}

Fig. \ref{fig:ccdf_2018} presents the CCDF for all flow types in 2018. This plot closely resembles the CCDF for 2010 shown in the main paper, indicating a remarkable stability in the strength distributions over time. The consistent tail behavior across both years suggests that the potential heavy-tailed relationships and distribution characteristics for different flow types remain largely unchanged. This similarity underscores the persistent nature of the network structure and flow patterns in the European regional system, with minimal alterations in the relative strengths of connections across various domains between 2010 and 2018.

\subsection{Pagerank}

\begin{figure}[t]
    \centering
    \includegraphics[width=\textwidth]{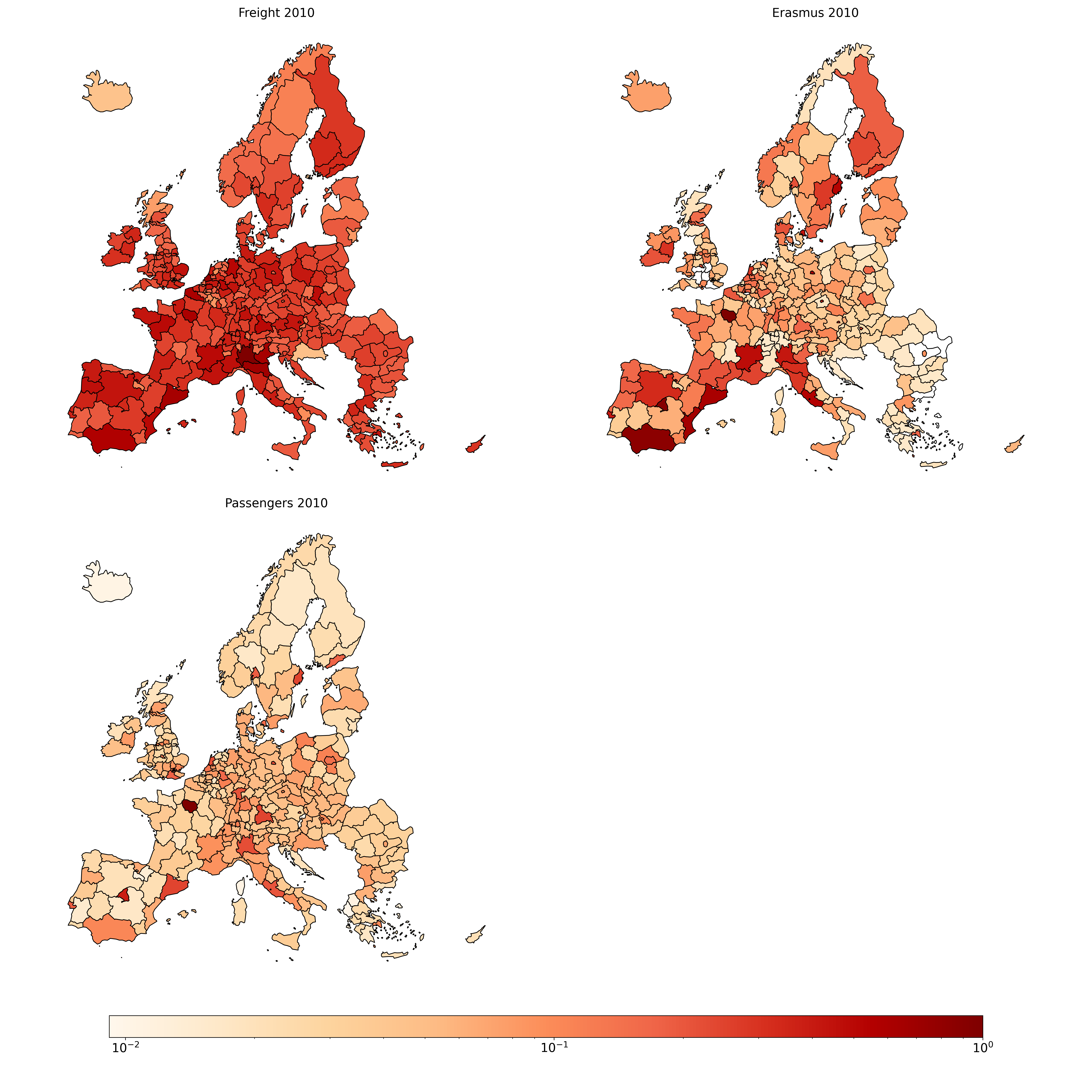}
    \caption{PageRank for Freight, Erasmus, and Passengers in 2010. Colors are displayed on a logarithmic scale, with values normalized such that the region with the highest centrality is set to 1.}
        \label{fig:3_layers}
\end{figure}

\begin{table}[h]
\centering
\begin{tabular}{lr}
\hline
Region & PageRank \\
\hline
Oberbayern & 1.0000 \\
Ile-de-France & 0.8087 \\
Inner London - East & 0.6598 \\
Darmstadt & 0.6441 \\
Stuttgart & 0.6259 \\
Düsseldorf & 0.6252 \\
Berlin & 0.6167 \\
Köln & 0.6105 \\
Attiki & 0.6045 \\
Lombardia & 0.6024 \\
\hline
\end{tabular}
\caption{Top 10 regions by PageRank for Migration in 2010.}
\label{tab:PageRank_migration_2010}
\end{table}

\begin{table}[h]
\centering
\begin{tabular}{lr}
\hline
Region & PageRank \\
\hline
Cataluña & 1.0000 \\
Ile-de-France & 0.9793 \\
Jadranska Hrvatska & 0.9237 \\
Andalucía & 0.7983 \\
Rhône-Alpes & 0.6389 \\
Provence-Alpes-Côte d’Azur & 0.6367 \\
Veneto & 0.5567 \\
Comunidad Valenciana & 0.5176 \\
Lombardia & 0.4910 \\
Comunidad de Madrid & 0.4865 \\
\hline
\end{tabular}
\caption{Top 10 regions by PageRank for Tourism in 2010.}
\label{tab:PageRank_tourism_2010}
\end{table}

\begin{table}[h]
\centering
\begin{tabular}{lr}
\hline
Region & PageRank \\
\hline
Lombardia & 1.0000 \\
Zuid-Holland & 0.7975 \\
Emilia-Romagna & 0.6890 \\
Veneto & 0.6428 \\
Cataluña & 0.6336 \\
Ile-de-France & 0.6021 \\
Nord-Pas de Calais & 0.5995 \\
Andalucía & 0.5737 \\
Weser-Ems & 0.5091 \\
Comunidad Valenciana & 0.5006 \\
\hline
\end{tabular}
\caption{Top 10 regions by PageRank for Freight in 2010.}
\label{tab:PageRank_freight_2010}
\end{table}

\begin{table}[h]
\centering
\begin{tabular}{lr}
\hline
Region & PageRank \\
\hline
Ile-de-France & 1.0000 \\
Comunidad de Madrid & 0.9202 \\
Andalucía & 0.8484 \\
Comunidad Valenciana & 0.6656 \\
Cataluña & 0.6186 \\
Berlin & 0.5657 \\
Lazio & 0.5479 \\
Stockholm & 0.5269 \\
Rhône-Alpes & 0.4794 \\
Hovedstaden & 0.4272 \\
\hline
\end{tabular}
\caption{Top 10 regions by PageRank for Erasmus in 2010.}
\label{tab:PageRank_erasmus_2010}
\end{table}

\begin{table}[h]
\centering
\begin{tabular}{lr}
\hline
Region & PageRank \\
\hline
Noord-Holland & 1.0000 \\
Ile-de-France & 0.6988 \\
Comunidad de Madrid & 0.5451 \\
Lombardia & 0.4559 \\
Bucureşti - Ilfov & 0.4537 \\
Luxembourg & 0.4333 \\
Eastern and Midland & 0.4220 \\
Région de Bruxelles-Capitale & 0.3791 \\
Zuid-Holland & 0.3750 \\
Warszawski stołeczny & 0.2924 \\
\hline
\end{tabular}
\caption{Top 10 regions by PageRank for FDI in 2010.}
\label{tab:PageRank_fdi_2010}
\end{table}

\begin{table}[h]
\centering
\begin{tabular}{lr}
\hline
Region & PageRank \\
\hline
Ile-de-France & 1.0000 \\
Comunidad de Madrid & 0.3822 \\
Inner London - West & 0.3775 \\
Inner London - East & 0.3187 \\
Stockholm & 0.2466 \\
Oberbayern & 0.2449 \\
Cataluña & 0.2446 \\
Berlin & 0.2232 \\
Noord-Holland & 0.2197 \\
Lombardia & 0.2186 \\
\hline
\end{tabular}
\caption{Top 10 regions by PageRank for Passengers in 2010.}
\label{tab:PageRank_passengers_2010}
\end{table}

\begin{table}[h]
\centering
\begin{tabular}{lr}
\hline
Region & PageRank \\
\hline
Ile-de-France & 1.0000 \\
Cataluña & 0.9165 \\
Luxembourg & 0.7257 \\
Comunidad de Madrid & 0.7015 \\
Vidurio ir vakarų Lietuvos regionas & 0.6988 \\
Prov. Antwerpen & 0.5824 \\
Rhône-Alpes & 0.5389 \\
Latvija & 0.4810 \\
Prov. Oost-Vlaanderen & 0.4793 \\
Comunidad Valenciana & 0.4563 \\
\hline
\end{tabular}
\caption{Top 10 regions by PageRank for Remittances in 2010.}
\label{tab:PageRank_remittances_2010}
\end{table}

Fig. \ref{fig:3_layers} illustrates the spatial distribution of PageRank centrality values across European regions for Freight, Erasmus, and Passengers in 2010.

Tables \ref{tab:PageRank_migration_2010} through \ref{tab:PageRank_remittances_2010} present the top 10 regions ranked by PageRank for various flow types in 2010.
For migration flows (Table \ref{tab:PageRank_migration_2010}), Oberbayern emerges as the most central region, followed closely by Ile-de-France and Inner London - East. The list is dominated by German regions, highlighting Germany's significance in European migration patterns.
Tourism flows (Table \ref{tab:PageRank_tourism_2010}) show Cataluña as the top-ranked region, with Ile-de-France and Jadranska Hrvatska following closely. This ranking reflects the popularity of Mediterranean coastal regions for tourism.
In freight transport (Table \ref{tab:PageRank_freight_2010}), Lombardia leads, followed by Zuid-Holland and Emilia-Romagna, underscoring the importance of industrial and port regions in goods movement.
For Erasmus student exchanges (Table \ref{tab:PageRank_erasmus_2010}), Ile-de-France ranks first, followed by Comunidad de Madrid and Andalucía, indicating the attractiveness of these regions for international students.
In Foreign Direct Investment (FDI) flows (Table \ref{tab:PageRank_fdi_2010}), Noord-Holland tops the list, with Ile-de-France and Comunidad de Madrid following, reflecting the financial importance of these regions.
Passenger flows (Table \ref{tab:PageRank_passengers_2010}) are dominated by Ile-de-France, with a significant lead over Comunidad de Madrid and Inner London - West, highlighting Paris's role as a major transportation hub.
Finally, for remittance flows (Table \ref{tab:PageRank_remittances_2010}), Ile-de-France again leads, followed by Cataluña and Luxembourg, indicating the economic significance of these regions for international money transfers.
These rankings collectively demonstrate the varied roles that different regions play in different types of flows, with some regions, particularly Ile-de-France, showing high centrality across multiple networks.

\begin{figure}[t]
    \centering
    \includegraphics[width=\textwidth]{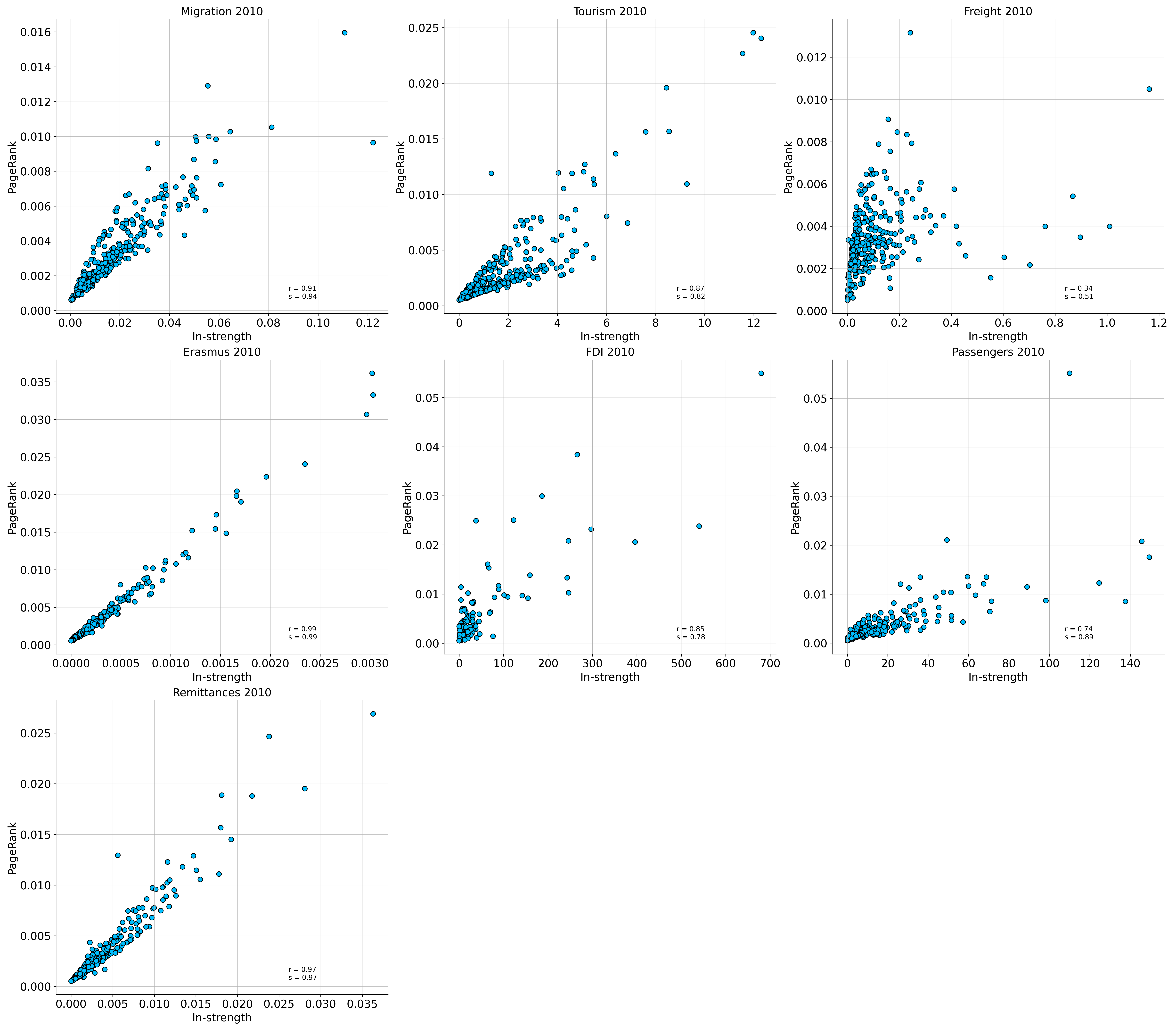}
    \caption{PageRank VS in-strength for the year 2010.}
        \label{fig:page_in_2010}
\end{figure}

Fig. \ref{fig:page_in_2010} demonstrates the relationship between PageRank and in-strength across all flow types in 2010. The analysis reveals strong correlations for most flow types, with both Pearson and Spearman correlation coefficients exceeding 0.74. This indicates a robust association between a region's centrality and the volume of incoming flows for most networks. However, the Freight network stands out as an exception, exhibiting a notably weaker correlation.

\begin{figure}[t]
    \centering
    \includegraphics[width=\textwidth]{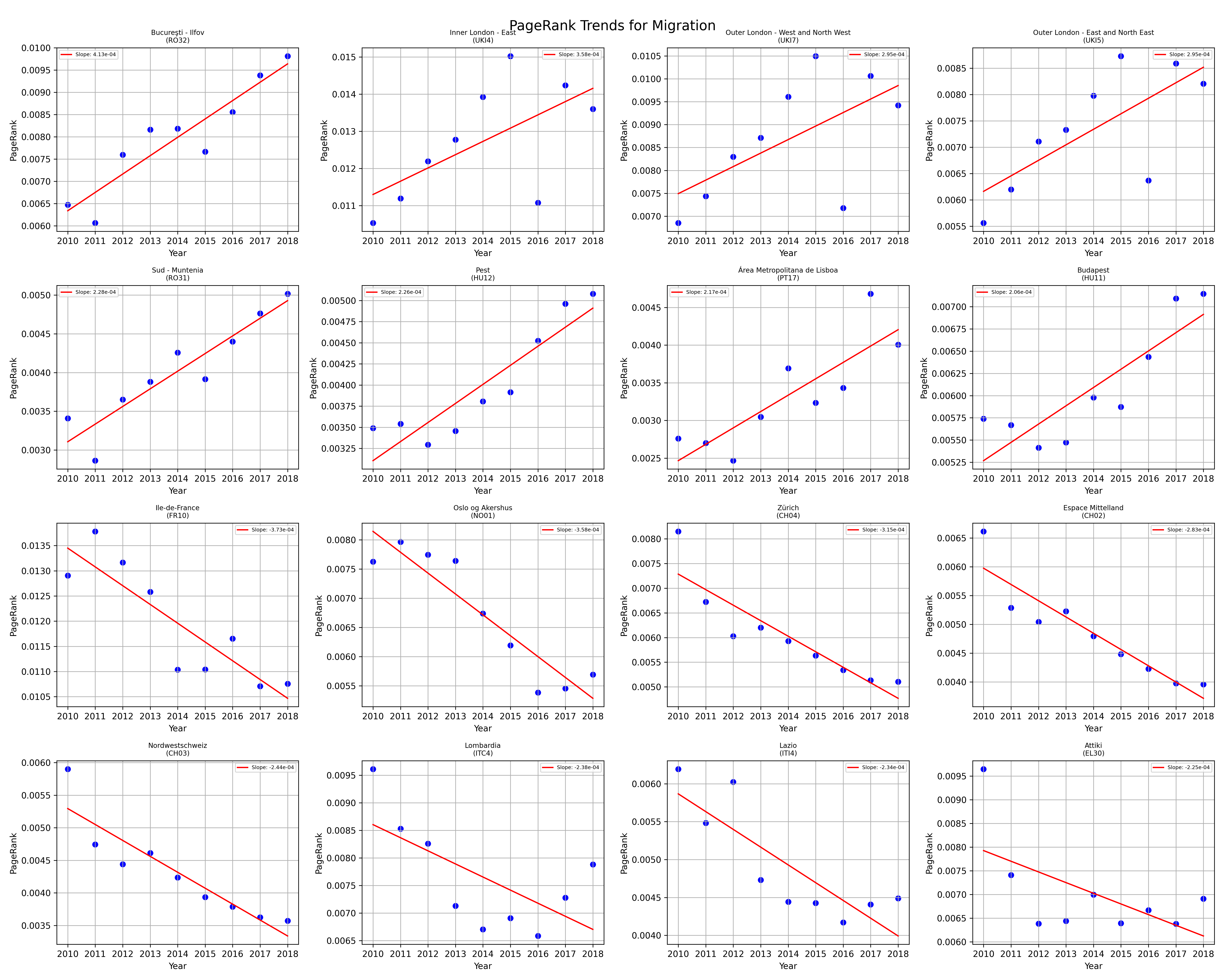}
    \caption{Top 8 and bottom 8 slopes of PageRank Trends for Migration.}
        \label{fig:page_slo}
\end{figure}

Fig. \ref{fig:page_slo} illustrates the top 8 and bottom 8 slopes of PageRank trends for Migration, highlighting regions with notably increasing or decreasing centrality. Interestingly, London stands out among the top increasing trends, despite experiencing a noticeable dip in 2016, likely attributable to the Brexit referendum. This overall upward trajectory, even in the face of such a significant political event, underscores London's resilience and enduring importance as a migration hub.

\begin{figure}[t]
    \centering
    \includegraphics[width=\textwidth]{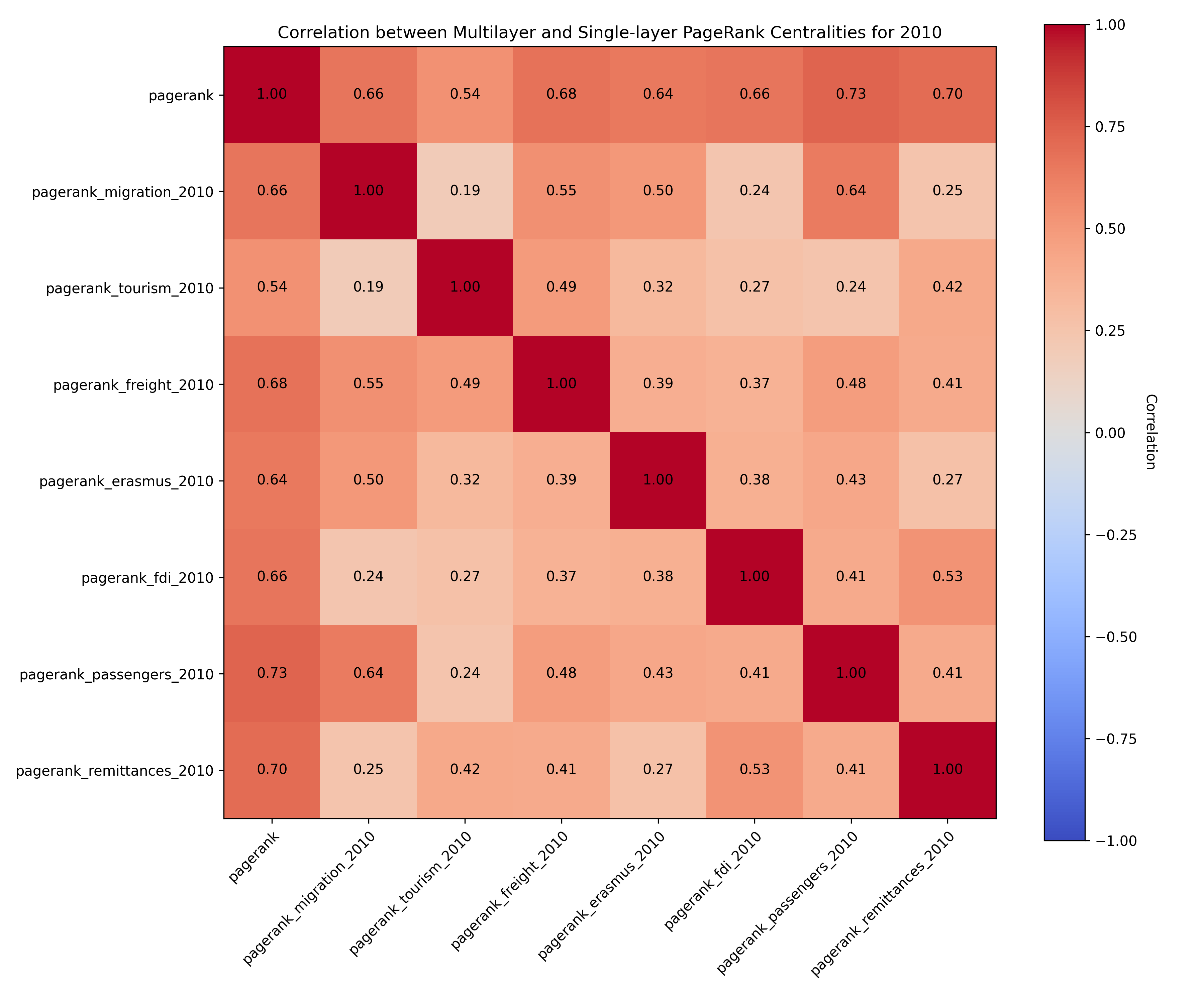}
    \caption{Correlations among single-layer PageRank and multiplex PageRank for 2010.}
        \label{fig:page_corr}
\end{figure}

Fig. \ref{fig:page_corr} displays the correlations between single-layer PageRank values and the multiplex PageRank for various flow types in 2010. The Spearman correlation coefficients range from 0.54 to 0.73, indicating moderate positive relationships between individual layer centralities and the overall multiplex centrality.


\subsection{Community Detection}
\begin{figure}[h]
    \centering
    \includegraphics[width=\textwidth]{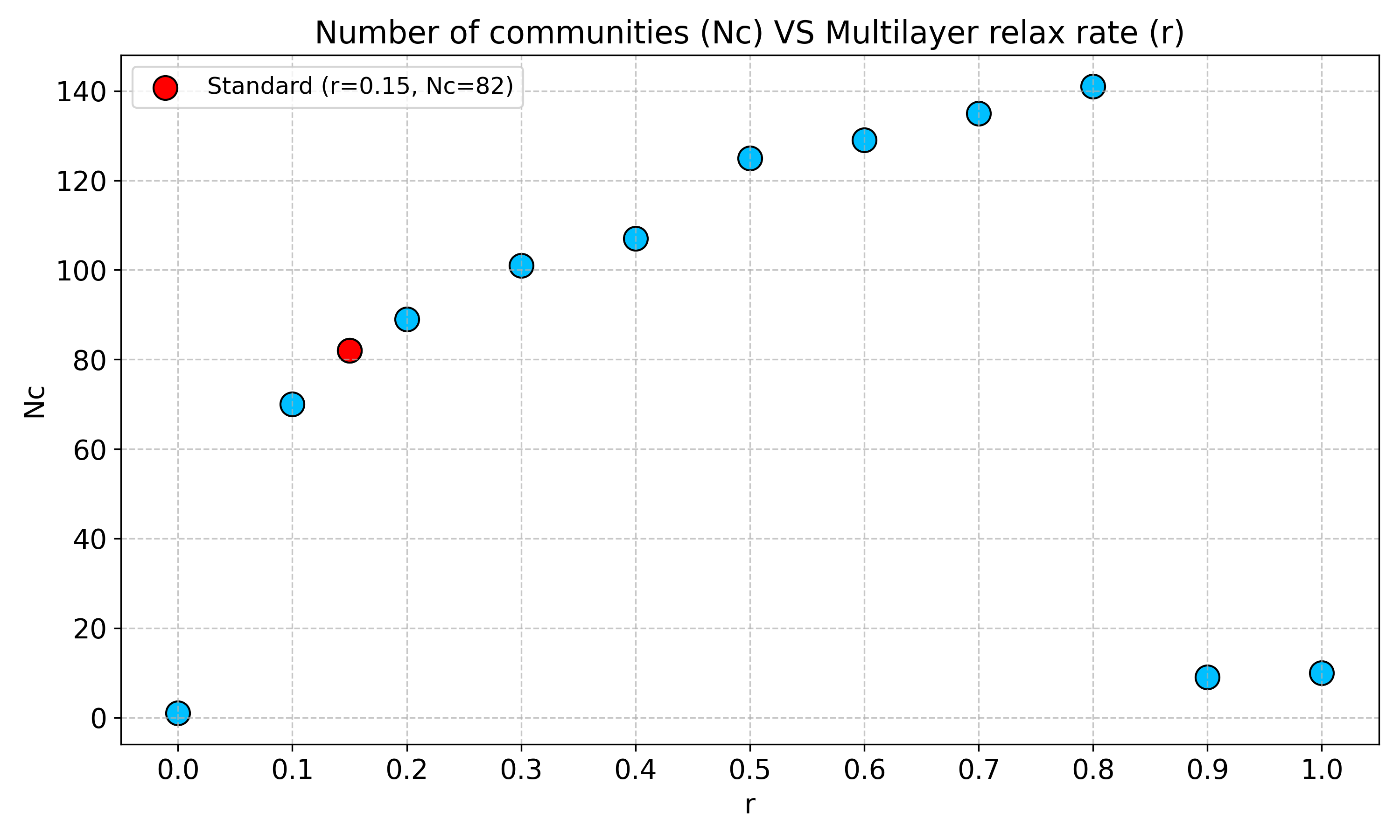} 
    \caption{Number of communities VS multilayer relax rate for 2010.}
    \label{fig:Nc}
\end{figure}

Fig. \ref{fig:Nc} illustrates the relationship between the number of communities and the multilayer relax rate for 2010, showing that the range around the standard value (from 0.1 to 0.2) yields a relatively stable number of communities, varying from 70 to 89. Generally, increasing the relax rate r leads to a higher number of communities, until it reaches an extremely high value (0.9), at which point the number of communities sharply decreases to around 10.

\begin{table}[h]
\caption{Infomap community detection results.}
\label{tab:comm_res}
\begin{adjustbox}{center}
\scriptsize
\begin{tabular}{|cc|cc|cc|cc|cc|cc|}
\hline
NUTS ID & Community & NUTS ID & Community & NUTS ID & Community & NUTS ID & Community & NUTS ID & Community & NUTS ID & Community \\
\hline
FRM0 & 1 & BE23 & 4 & CY00 & 8 & HU23 & 14 & DE11 & 23 & DE93 & 40 \\
FR10 & 1 & BE22 & 4 & UKK1 & 8 & HU31 & 14 & DE12 & 23 & UKK4 & 41 \\
FRL0 & 1 & BE21 & 4 & UKJ4 & 8 & HU33 & 14 & DE14 & 23 & UKK3 & 41 \\
FRE2 & 1 & BE10 & 4 & UKJ2 & 8 & HU22 & 14 & DE13 & 23 & DE24 & 42 \\
FRY3 & 1 & NL34 & 4 & UKJ3 & 8 & HU32 & 14 & UKN0 & 24 & DE25 & 42 \\
FRB0 & 1 & BE25 & 4 & RO11 & 9 & HU12 & 14 & IE04 & 24 & DE94 & 43 \\
FRK2 & 1 & LU00 & 4 & RO41 & 9 & HU11 & 14 & IE05 & 24 & DED2 & 44 \\
FRJ1 & 1 & ITC2 & 5 & RO21 & 9 & HU21 & 14 & IE06 & 24 & DEG0 & 45 \\
FRI3 & 1 & ITC1 & 5 & RO32 & 9 & DK04 & 15 & UKE2 & 25 & FRF1 & 46 \\
FRI2 & 1 & ITH3 & 5 & RO31 & 9 & DK02 & 15 & UKE3 & 25 & FRY1 & 47 \\
FRI1 & 1 & ITC4 & 5 & RO42 & 9 & DK03 & 15 & UKE4 & 25 & FRY2 & 47 \\
FRH0 & 1 & ITI4 & 5 & RO12 & 9 & DK01 & 15 & UKC2 & 25 & DED4 & 48 \\
FRJ2 & 1 & ITF5 & 5 & RO22 & 9 & DK05 & 15 & UKC1 & 25 & FRF3 & 49 \\
FRG0 & 1 & ITF6 & 5 & AT11 & 10 & HR03 & 16 & UKE1 & 25 & ES62 & 50 \\
ES70 & 2 & ITF2 & 5 & AT34 & 10 & MT00 & 16 & DE60 & 26 & FRD2 & 51 \\
ES64 & 2 & ITI1 & 5 & AT33 & 10 & SI03 & 16 & DE80 & 26 & FRF2 & 52 \\
ES63 & 2 & ITH5 & 5 & AT12 & 10 & SI04 & 16 & DEF0 & 26 & ITF3 & 53 \\
ES61 & 2 & NL32 & 6 & AT13 & 10 & HR04 & 16 & DE27 & 27 & DE73 & 54 \\
ES53 & 2 & NL11 & 6 & AT21 & 10 & NO06 & 17 & DE21 & 27 & DEA4 & 55 \\
ES51 & 2 & NL12 & 6 & AT22 & 10 & NO02 & 17 & UKD1 & 28 & ITC3 & 56 \\
ES42 & 2 & NL13 & 6 & AT32 & 10 & NO03 & 17 & UKD6 & 28 & FRC1 & 57 \\
ES23 & 2 & NL21 & 6 & AT31 & 10 & NO05 & 17 & UKD3 & 28 & DEC0 & 58 \\
ES24 & 2 & NL33 & 6 & SE23 & 11 & NO07 & 17 & UKD4 & 28 & FRY4 & 59 \\
ES30 & 2 & NL42 & 6 & SE33 & 11 & NO04 & 17 & UKD7 & 28 & FRY5 & 59 \\
ES41 & 2 & NL31 & 6 & SE21 & 11 & NO01 & 17 & UKM9 & 29 & ITF4 & 60 \\
ES11 & 2 & NL23 & 6 & SE22 & 11 & EE00 & 18 & UKM5 & 29 & DE23 & 61 \\
ES52 & 2 & NL22 & 6 & SE12 & 11 & LV00 & 18 & UKM6 & 29 & DE72 & 62 \\
PL62 & 3 & NL41 & 6 & SE11 & 11 & LT02 & 18 & UKM7 & 29 & DE26 & 63 \\
PL51 & 3 & CZ06 & 7 & SE32 & 11 & LT01 & 18 & UKM8 & 29 & FRC2 & 64 \\
PL52 & 3 & SK04 & 7 & FI20 & 11 & CH04 & 19 & DE30 & 30 & DE22 & 65 \\
PL61 & 3 & CZ02 & 7 & SE31 & 11 & LI00 & 19 & DE40 & 30 & ITG1 & 66 \\
PL63 & 3 & CZ04 & 7 & EL65 & 12 & CH03 & 19 & UKG3 & 31 & ITI3 & 67 \\
PL71 & 3 & CZ03 & 7 & EL53 & 12 & CH05 & 19 & UKG1 & 31 & ITH2 & 68 \\
PL72 & 3 & CZ01 & 7 & EL52 & 12 & CH01 & 19 & UKG2 & 31 & ES22 & 69 \\
PL81 & 3 & CZ07 & 7 & EL51 & 12 & CH02 & 19 & UKF1 & 32 & ITI2 & 70 \\
PL92 & 3 & SK01 & 7 & EL54 & 12 & CH06 & 19 & UKF2 & 32 & DEB2 & 71 \\
PL91 & 3 & SK02 & 7 & EL42 & 12 & BG32 & 20 & UKF3 & 32 & ITH1 & 72 \\
PL82 & 3 & SK03 & 7 & EL41 & 12 & BG31 & 20 & DE91 & 33 & ITH4 & 73 \\
PL43 & 3 & CZ05 & 7 & EL30 & 12 & BG42 & 20 & DE92 & 33 & ITG2 & 74 \\
PL41 & 3 & CZ08 & 7 & EL61 & 12 & BG33 & 20 & DE71 & 34 & FRK1 & 75 \\
PL42 & 3 & UKH3 & 8 & EL62 & 12 & BG34 & 20 & DEB1 & 35 & PT15 & 76 \\
PL84 & 3 & UKJ1 & 8 & EL63 & 12 & BG41 & 20 & DEB3 & 35 & FRD1 & 77 \\
PL21 & 3 & UKK2 & 8 & EL64 & 12 & DEA2 & 21 & DED5 & 36 & ITF1 & 78 \\
PL22 & 3 & UKH1 & 8 & EL43 & 12 & DEA3 & 21 & DEE0 & 36 & ES43 & 79 \\
BE35 & 4 & UKH2 & 8 & PT30 & 13 & DEA5 & 21 & ES12 & 37 & ES13 & 80 \\
BE34 & 4 & UKI3 & 8 & PT11 & 13 & DEA1 & 21 & ES21 & 37 & CH07 & 81 \\
BE33 & 4 & UKI4 & 8 & PT16 & 13 & FI1D & 22 & UKL2 & 38 & IS00 & 82 \\
BE32 & 4 & UKI5 & 8 & PT17 & 13 & FI1C & 22 & UKL1 & 38 &  &  \\
BE31 & 4 & UKI6 & 8 & PT18 & 13 & FI1B & 22 & FRE1 & 39 &  &  \\
BE24 & 4 & UKI7 & 8 & PT20 & 13 & FI19 & 22 & DE50 & 40 &  &  \\
\hline
\end{tabular}
\end{adjustbox}
\end{table}

Table \ref{tab:comm_res} presents the results of the Infomap community detection analysis.

\end{document}